\pdfoutput=1
\documentclass[aps,preprintnumbers,showpacs,twocolumn,superscriptaddress,floatfix,nofootinbib]{revtex4-1}

\usepackage{latexsym} 
\usepackage{amssymb} 
\usepackage{amsmath} 
\usepackage{amsfonts}
\usepackage{bm}
\usepackage{color}
\usepackage{times} 
\usepackage{units}
\usepackage{hyperref}

\usepackage[utf8x]{inputenc}

\usepackage{graphicx} 
\usepackage[squaren]{SIunits} 
\usepackage[inline]{enumitem}

\newcommand{\Eref}[1]{Eq.~\eqref{#1}} 
\newcommand{\Fref}[1]{Fig.~\ref{#1}} 
\newcommand{\Sref}[1]{Sec.~\ref{#1}} 
\newcommand{\Tref}[1]{Table~\ref{#1}}

\newcommand{\Fced}{\bar{\tau}}
\newcommand{\Fcrmd}{\bar{D}}
\newcommand{\Fcmom}{\bar{S}}
\newcommand{\Fq}{\bar{q}}
\newcommand{\Fr}{\bar{r}}

\newcommand{\Tced}{\tau}
\newcommand{\Tcrmd}{D}
\newcommand{\Tcmom}{S}
\newcommand{\Tq}{q}
\newcommand{\Tr}{r}
\newcommand{\Trp}{r_\|}
\newcommand{\Tro}{r_\bot}
\newcommand{\Tb}{b}
\newcommand{\Tx}{x}
\newcommand{\Tmu}{\mu}

\begin{document}

\title{Robust Recovery of Primitive Variables in Relativistic Ideal Magnetohydrodynamics}

\author{Wolfgang Kastaun}
\affiliation{Max Planck Institute for Gravitational Physics (Albert
  Einstein Institute), Callinstrasse 38, 30167 Hannover, Germany}
\affiliation{Leibniz Universit\"at Hannover, 30167 Hannover, Germany}

\author{Jay Vijay Kalinani}
\affiliation{Universit\`a di Padova, Dipartimento di Fisica e Astronomia, Via Francesco Marzolo 8, I-35131 Padova, Italy}
\affiliation{INFN, Sezione di Padova, Via Francesco Marzolo 8, I-35131 Padova, Italy}

\author{Riccardo Ciolfi}
\affiliation{INAF, Osservatorio Astronomico di Padova, Vicolo dell'Osservatorio 5, I-35122 Padova, Italy}
\affiliation{INFN, Sezione di Padova, Via Francesco Marzolo 8, I-35131 Padova, Italy}

\begin{abstract}
\noindent Modern simulation codes for general relativistic ideal magnetohydrodynamics  
are all facing
a long standing technical problem given by the need to recover fundamental
variables from those variables that are evolved in time. In the relativistic case,
this requires the numerical solution of a system of nonlinear equations. Although 
several approaches are available, none has proven completely reliable. A recent 
study comparing different methods showed that all can fail, in particular for 
the important case of strong magnetization and moderate Lorentz factors.
Here, we propose a new robust, efficient, and accurate solution scheme, 
along with a proof for the existence and uniqueness of a solution, and analytic
bounds for the accuracy. Further, the scheme allows us to reliably detect
evolution errors leading to unphysical states and automatically applies  
corrections for typical harmless cases. A reference implementation
of the method is made publicly available as a software library. The aim 
of this library is to improve the reliability of 
binary neutron star merger simulations, in particular in the investigation of 
jet formation and magnetically driven winds.
\end{abstract}


\maketitle

%===========================================================================
\section{Introduction}
\label{sec:intro}

\noindent General relativistic magnetohydrodynamic (GRMHD) simulations are an important
tool to study many astrophysical scenarios involving neutron stars (NSs) and black holes (BHs).
In particular, they represent the leading approach to investigate the dynamics of binary neutron star (BNS) and NS-BH mergers, which are the most important events in the nascent field of multimessenger astrophysics with gravitational wave (GW) sources \cite{LVC:MMA:2017}.

Arguably one of the most pressing unsolved problems related to BNS and NS-BH mergers is to find the exact mechanism powering short gamma-ray bursts (SGRBs).  
The simultaneous detection of the gravitational wave event GW170817 and the SGRB named GRB\,170817A \cite{LVC:BNSDetection,LVC:BNSSourceProp:2019,LVC:GWGRB:2017}, along with the following ``afterglow'' emission across the entire electromagnetic spectrum (e.g., \cite{LVC:MMA:2017,Troja2017,Hallinan2017,Lyman2018}), provided compelling evidence that BNS mergers can power SGRBs (e.g., \cite{Lazzati2018,Mooley2018b,Ghirlanda2019}).
In turn, this implies that the remnant object resulting from a BNS merger can launch, at least in some cases, a relativistic jet, which was indeed confirmed for GRB\,170817A \cite{Mooley2018b,Ghirlanda2019}. 
However, the jet launching mechanism and the nature of the object acting as a central engine, either an accreting BH or a massive NS, remain uncertain (e.g., \cite{Ciolfi2018}).

Current simulations suggest that a mechanism based on neutrino-antineutrino annihilation would not be powerful enough to explain SGRBs \cite{Just2016,Perego2017a}, reinforcing the alternative idea that the main driver of jet formation should be a strong magnetic field. 
GRMHD simulations of BNS and NS-BH mergers, while considerably more complex and expensive because of the inclusion of magnetic fields, become necessary to properly address the problem.
Recent studies in this direction already provided important hints, supporting a scenario where the central engine is an accreting BH \cite{Paschalidis:2015:14,Ruiz2016} while disfavoring the massive NS scenario \cite{Ciolfi2020a}. Nonetheless, a final answer is still missing, and it will be necessary to overcome the technical limitations of present GRMHD codes to ultimately solve the problem. 

The merger event GW170817 was also accompanied by the kilonova transient AT\,2017gfo, powered by the radioactive decay of heavy r-process elements synthesized within the matter ejected by the merger (e.g., \cite{LVC:MMA:2017,Pian2017}).
Although this kilonova was observed in unprecedented detail, the interpretation in terms of specific ejecta components and their physical origin is still under debate. 
Also in this case, numerical relativity simulations represent the ideal approach to fully understand the different mass ejection processes occurring in a BNS (or a NS-BH) merger. Moreover, for some of these ejection processes magnetic fields are likely to play an important role (e.g., \cite{Siegel2018,CiolfiKalinani2020}) and therefore simulations should be performed in GRMHD.

The present work is devoted to a technical but crucial aspect of these simulations that has proven surprisingly difficult, 
and is motivated by the importance of GRMHD simulations in the context of BNS mergers (see, e.g., \cite{Ciolfi2020b} for a recent review).
Modern evolution codes are based on 
evolution equations written in form of coupled conservation laws for baryon number 
density, energy and momentum density including the electromagnetic contributions, 
and either magnetic field or vector potential. 
Primitive variables such as matter velocity, density, and pressure,
are not directly evolved. Instead, they have to be recovered from the evolved
quasi-conserved quantities after each evolution step. 

While in Newtonian physics the above recovery is trivial, for the relativistic case one has to numerically solve a system of coupled nonlinear equations. 
The system involves also the equation of state (EOS), which encodes the behavior of matter up to supranuclear densities by specifying the pressure as a function of density and temperature. 
An additional degree of freedom is the electron fraction, which effectively
describes the matter composition and which can only change due to neutrino reactions. 
Since the EOS is not well constrained theoretically or by observation,
a crucial requirement is the ability to perform simulations 
employing arbitrary EOS.
This precludes closed-form solutions for the primitive variables, 
and the system has to be solved numerically.
Since the solution is required inside the innermost loop of the evolution,
computational efficiency is almost as important as robustness.

Note that most evolution codes make the simplifying assumption of ideal MHD.
Although the electrical conductivity in merger remnants is very high, this approximation might not be justified for all aspects of the problem. On the other hand, evolving resistive GRMHD equations introduces even more difficulties (see also \cite{Wright:2019:stz2779}). The equations for the primitive variable recovery are
also very different for resistive GRMHD.
Another complication is that in regions with strong magnetic fields but low 
mass density, movement of the matter becomes dominated
by the field. Treating this ``force-free'' regime would in principle require different 
numerical evolution methods (for example, see \cite{Etienne:2017:215001}).

The simpler problem of recovering the primitive variables in relativistic 
hydrodynamics without magnetic fields is already solved in a robust manner, 
as described in \cite{Galeazzi:2013:64009} (also adopted
in \cite{Radice:2012:2012}). 
For the full problem of ideal GRMHD, several recovery methods with different
limitations have been employed in GRMHD evolution codes \cite{Giacomazzo:2007:235,Duran:2008:937,Moesta:2014:015005,Etienne:2015:175009,
Cipolletta:2019:arXiv,Noble:2006:626,Neilsen:2014:104029,Palenzuela:2015:044045}. 
Older schemes such as \cite{Noble:2006:626} are limited to particular analytic 
prescriptions for the EOS. Newer schemes can in principle work
with any EOS, but not all implementations actually allow the use of arbitrary EOS.
For a detailed review, we refer to \cite{Siegel:2018:71}.

All of the schemes investigated in \cite{Siegel:2018:71} were shown to fail in 
certain regimes. 
While some of them work well enough in most of the regimes encountered during a merger simulation, even rare primitive recovery failures need to be handled and remain a common hurdle.
An additional complication is that not all combinations of values for the evolved
variables correspond to physically valid primitive variables. The occurrence
of invalid evolved variables due to numerical errors of the evolution needs to
be monitored and, if possible, corrected. If the recovery can fail also for 
valid input, it becomes impossible to reliably assess the overall validity.

In this work, we developed a new recovery algorithm with the mathematically proven
ability to always find a solution, and which is guaranteed to recognize invalid evolved variables.
Furthermore, the scheme provides mathematically
derived accuracy bounds. Our scheme is limited to the ideal MHD approximation,
but it does not introduce problems in the force-free regime.
We provide a reference implementation which is ready to use in any GRMHD 
evolution code, in the form of a C++ library named \texttt{RePrimAnd} \cite{RePrimAnd}. 
Our implementation is written to be completely EOS-agnostic and provides 
a framework for EOS that can easily be extended.
Since our aim is to improve reliability, we subject the numerical implementation
of the algorithm to a comprehensive suite of tests, also studying the effects of finite
floating point precision.

The article is organized as follows: In \Sref{sec:formulation}, we state the problem,
derive the new scheme, prove the existence and uniqueness of a 
solution, and investigate the expected accuracy.
In \Sref{sec:enforce}, we discuss possible corrections to invalid evolved variables.
In \Sref{sec:perf}, we present numerical tests of our reference implementation, demonstrating 
robustness, efficiency, and precision. Here we also compare to other existing schemes.
Then, we study error propagation of evolution errors to the primitive variables
in \Sref{sec:evol_err}, identifying potentially problematic regimes. Finally, in \Sref{sec:summary} we draw our conclusions. 

%===========================================================================
\section{Formulation of the Scheme}
\label{sec:formulation}
\subsection{Primitive variables}
\label{sec:primvars}

Our scheme is designed for evolution codes which evolve 
variables defined on a spacelike foliation of spacetime from 
one timeslice to the next. The hypersurfaces and their normal observers 
define a frame we will refer to as the Eulerian frame. 

We denote the 3-velocity of the fluid with respect to the Eulerian
frame as $v^i$, and the corresponding Lorentz factor as $W$. 
We will also use a quantity $z\equiv Wv$. 
The baryon number density in the fluid restframe is denoted as $n_\mathrm{B}$.
It is common to multiply $n_\mathrm{B}$ with an arbitrary mass 
constant $m_\mathrm{B}$ to define the baryonic mass density 
$\rho = n_\mathrm{B} m_\mathrm{B}$.
The pressure in the fluid restframe is assumed to be isotropic 
and denoted as $P$. 
Denoting the fluid contribution to the total energy density in the 
fluid restframe as $\rho_E$, we define the specific internal energy
\begin{align}
\epsilon &= \frac{\rho_E}{\rho} - 1
\end{align}
We further define $a=P/\rho_E$ and the relativistic enthalpy
\begin{align}
h &= 1 + \epsilon + \frac{P}{\rho} 
= \left( 1 + \epsilon \right) \left( 1 + a \right) \label{eq:defenthalpy}
\end{align}
Note that the definitions of $\epsilon$ and $h$ both depend on the 
arbitrary choice of $m_\mathrm{B}$. 

The primitive variables we use to describe the electromagnetic field are
the electric and magnetic fields as seen by an Eulerian observer. In terms
of the Maxwell tensor,
\begin{align}
E^\mu &= n_\nu F^{\mu\nu}, &
B^\mu &= n_\nu {}^*F^{\mu\nu}, &
\end{align}
where $n$ is the normal to the hypersurfaces of the foliation, and 
the star denotes the Hodge dual. $E^\mu, B^\mu$ are tangential to the 
hypersurface and thus equivalent to 3-vectors $E^i, B^i$. 
Our scheme neither requires nor provides the fields in the fluid frame,
which can be obtained from the above using standard transformations.

\subsection{Equation of State}
\label{sec:eos}
We assume an equation of state (EOS) of the form
\begin{align}
P &= P(\rho, \epsilon)
\end{align}
The EOS could also depend on further variables, such as the electron fraction,
as long as those variables are evolved variables or can be obtained from evolved
variables in a trivial way, and can therefore be treated as fixed parameters
in the primitive recovery algorithm.

For our purpose, it is also important to specify a validity range for each EOS.
The validity range considers both physical and technical constraints.
The most important physical constraint is the zero temperature limit for the 
internal energy. An example of a technical constraint is the range of values
available for an EOS given in tabulated form. Currently, our scheme uses
an EOS-dependent validity region specified in the following form
\begin{align}
\rho_\text{min} &\le \rho \le \rho_\text{max} \\
\epsilon_\text{min}(\rho) &\le \epsilon \le \epsilon_\text{max}(\rho) 
\end{align}
However, it could easily be adapted to a more general shape in $\rho,\epsilon$ 
parameter space.
We require that the lower validity bound $\epsilon_\text{min}(\rho)$ be
the zero-temperature value at the given density.
This is the only meaningful choice for any numerical simulation involving cold 
matter at any time.
Our error policy for correcting invalid evolved variables is based on 
this assumption, as is the proof for guaranteed success of the algorithm.

Our scheme relies on some physical constraints. 
Causality and thermodynamic stability require 
\begin{align}
0\le c_s^2 < 1 \, ,
\end{align}  
where $c_s$ is the adiabatic speed of sound, given by
\begin{align}\label{eq:csound}
c_s^2 &=
\left. \frac{\mathrm{d} \ln(h)}{\mathrm{d} \ln(\rho)} \right|_{s=\text{const}}  
\end{align}
If the EOS depends on electron fraction, it is also assumed to be constant in the 
above expression.
We assume positive baryon number density and positive total energy 
density, which implies
\begin{align}
0 &\le \rho_\text{min} \le \rho \, , &
-1 &< \epsilon_\text{min}(\rho) \le \epsilon 
\end{align}

We assume that the pressure is positive and further bounded
by the total energy density (dominant energy condition), which implies
\begin{align}
0 &\le a \le 1 
\end{align}
For a given EOS, we also require a positive lower bound $h_0$ for the relativistic enthalpy $h$,
such that $0<h_0\le h(\rho,\epsilon)$ over the entire validity region of the EOS. 
This requirement only excludes exotic matter with $P \le - \rho_E$.

Note that we do not assume $\epsilon>0$ or $h \ge 1$. The definitions of $\epsilon$ 
and $h$ depend on the arbitrary choice of the mass constant $m_B$. 
Unless the latter is fine-tuned to the average baryon binding energy at low density,
nuclear physics EOS often yield slightly negative $\epsilon$. 
In practice, $h_0$ is of order unity.

By design, our scheme is not confined to any particular equation of state. 
It only uses the information defined above and does not make any other 
kind of EOS-specific distinctions or adjustments.
For the purpose of testing our scheme, we use two specific EOS as examples:
\begin{enumerate}
\item A hybrid EOS given by a cold part and a simple thermal part
\begin{align}
P(\rho,\epsilon) &= P_\text{cold}(\rho) + 
\left(\Gamma_\text{th} -1\right) \rho \left( \epsilon - \epsilon_\text{cold}(\rho) \right) \\
\epsilon_\text{min}(\rho) &= \epsilon_\text{cold}(\rho)
\end{align}
For the cold part, we employ the MS1 EOS from \cite{Read:2009:124032} (based on \cite{Mueller1996}), and we use
$\Gamma_\text{th} = 1.8$. 
This hybridized type of EOS is often used in numerical relativity.
\item The classical ideal gas, given by 
\begin{align}
P(\rho,\epsilon) &= \rho\epsilon\left(\Gamma -1\right) \\
\epsilon_\text{min}(\rho) &= 0
\end{align}
Here, we use an adiabatic exponent $\Gamma = 2$.
Pressure and internal energy are zero at zero temperature. 
We use this unrealistic model, where pressure is only given by thermal 
effects and degeneracy pressure is ignored, as a corner case for testing 
our algorithm. 
Note that in numerical relativity, ideal gas EOS normally refers to 
a hybrid EOS based on a polytrope with adiabatic exponent 
$\Gamma = \Gamma_{th}$ as the zero-temperature limit.

\end{enumerate}

We note that numerical relativity tools are often judged by how well they 
function with tabulated nuclear physics EOS, because 
such tables often contain isolated points 
with thermodynamically inconsistent jumps or regimes with superluminal soundspeed. 
Violations of basic physical constraints such as causality can lead to many 
fundamental problems with the evolution equations. In \Sref{sec:uniqe}, we 
will point out a potential problem for primitive recovery. In our opinion, 
any effort to ensure that primitive variable recovery and other aspects of 
simulations can produce results with faulty EOS is a step in the wrong direction.
Instead, we advocate in favor of repairing 
minor EOS faults before use.

\subsection{Evolved Variables}
Our scheme is intended for numerical evolution codes employing 
evolution equations for energy, momentum, and baryon number density
formulated as a quasi-conservation law. 
This is the case for finite-volume shock-capturing schemes.
The evolved quantities are called conserved variables, although only 
the baryon number is strictly conserved.
The fluid contribution is given by
\begin{align}
\Fcrmd    &= \rho W \label{eq:def_dens}\\ 
\Fced     &= \Fcrmd \left( h W - 1 \right)  - P  \\
\Fcmom_i  &= \Fcrmd W h v_i
\end{align}
Including also EM contributions, the evolved variables are given by
\begin{align}
\Tcrmd    &= \Fcrmd \\
\Tced     &= \Fced + \frac{1}{2} \left( E^2 + B^2 \right) \\
\Tcmom_i  &= \Fcmom_i + \epsilon_{ijk} E^j B^k \label{eq:def_cmom_tot}
\end{align}
In most formulations, the evolved variables above include the volume element 
of the spatial metric. Since this factor is available from the spacetime 
evolution, it is not relevant for the following and was left out of 
the definitions.

In addition to the evolved variables $\Tcrmd, \Tced, \Tcmom_i$, we assume 
that the magnetic field $B$ in the Eulerian frame is either an evolved variable
or can be reconstructed from evolved variables without knowledge of the 
fluid-related primitive variables, such that $B$ is known when our primitive
reconstruction scheme is run. This is the case for state of the art methods such
as constrained transport schemes or schemes evolving the vector potential.

We only consider evolution codes that assume the ideal MHD limit,
which corresponds to the additional condition
\begin{align} \label{eq:idealmhd}
E^l &= -\epsilon^{ljk} v_j B_k 
\end{align}
This precludes the use of our scheme for resistive MHD simulations.

For convenience, we define rescaled variables as
\begin{align}
\Fq   &= \frac{\Fced}{\Fcrmd} &
\Fr_i &= \frac{\Fcmom_i}{\Fcrmd} 
\label{eq:defq}\\
\Tq   &= \frac{\Tced}{\Fcrmd} &
\Tr_i &= \frac{\Tcmom_i}{\Fcrmd} 
\\
\Tb_i &= \frac{B_i}{\sqrt{\Fcrmd}} &
\Tb &=\sqrt{ b^i b_i }
\label{eq:defbsqr}
\end{align}
It is worth noting that for most aspects of our scheme, the relevant scale 
for the magnetic field is given by $b$, not by the commonly used magnetization, 
defined as the ratio between magnetic and fluid pressure. Since the pressure
is typically orders of magnitude below the mass energy density, $\mathcal{O}(b)=1$ 
corresponds to a very large magnetization.

We also need to decompose the momentum into parts parallel and normal
to the magnetic field
\begin{align}
\Trp^i &= \frac{\Tb^l \Tr_l}{\Tb^2} \Tb^i &
\Tro^i &= \Tr^i - \Trp^i 
\end{align}
The decomposition is undefined for the case of zero magnetic field, but
we exclusively use the product with $b^2$ in our scheme, which is always well-defined.

\subsection{Useful Relations}
In the following, we collect definitions and analytic relations for later use.
First, we define two quantities that will play a central role,  
\begin{align}
\Tmu &\equiv \frac{1}{Wh} \, , &
\Tx     &\equiv \frac{1}{1 + \Tmu \Tb^2} \label{eq:defmuandx}
\end{align}
Trivially, their ranges are limited to 
\begin{align}
0 &<\Tmu \le 1/h_0 \, , &
0 &<\Tx \le 1
\end{align}

Given the conserved variables and $\Tmu$, one can directly 
express the primitive variables analytically. Since the system is 
over-determined, there are different possible expressions which may
disagree if the given $\Tmu$ is inconsistent with the conserved variables.
In the latter case, some expressions can diverge.
We will use the ambiguity to cast the recovery into a root-finding problem, 
by expressing the same variables in different ways that only agree 
for the correct value of $\mu$.

As an intermediate step, we first remove the electromagnetic part of the
conserved variables.
From \Eref{eq:idealmhd}
\begin{align}
E^2 &= \Tx^2 \Tmu^2 B^2 \Tro^2 \\
\Fr^i &= x \Tro^i + \Trp^i \label{eq:rfvec}
\end{align}
This allows us to compute the pure fluid part of the conserved variables.
The corresponding quantities relevant for our purpose can be written as
\begin{align}
\Fr^2 &= x^2 \Tro^2 + \Trp^2  \label{eq:rbsqr}\\
\Fq &= \Tq - \frac{1}{2} \Tb^2 - \frac{1}{2} \Tmu^2 x^2 \Tb^2 \Tro^2 \label{eq:qb}
\end{align}
We can now express the velocity as $v = \Tmu \Fr$. 
This expression does however not guarantee that $v<1$ for any $\Tmu$.
One way to avoid exceeding the speed of light is to use the quantity $z$,
which yields
\begin{align}
v(z) &= \frac{z}{\sqrt{1+z^2}} < 1
\end{align}
Although we do not have a closed form expression for $z$ as a
function of $\mu$, we can use $z$ to obtain a useful upper limit for the velocity,
given by
\begin{align}
z &= \frac{\bar{r}}{h} \le \frac{r}{h} \le \frac{r}{h_0} \equiv z_0,&
v \le v_0 \equiv v(z_0) \label{eq:vmax}
\end{align}
After obtaining the velocity and Lorentz factor, we can extract the 
rest mass density and specific internal energy using the expressions
\begin{align} 
\rho &= \bar{D} / W \label{eq:rhoraw} \\
\epsilon  &= W \left( \bar{q}  - \Tmu \bar{r}^2 \right) + W - 1
\label{eq:epsraw}
\end{align}
If $\rho$ and  $\epsilon$ are in the validity range of the EOS,
we can now compute the pressure $P = P(\rho, \epsilon)$ and 
the enthalpy $h(\rho, \epsilon)$.
Finally, the following expression for $\Tmu$ itself turns out to be 
useful
\begin{align}
\Tmu &= \frac{1}{hW} = \frac{1}{hW \left(W^{-2} + v^2\right)} \\
&=\frac{1}{\frac{h}{W} + \frac{v^2}{\Tmu}}
= \frac{1}{\frac{h}{W} + \bar{r}^2 \Tmu} \label{eq:mualt}
\end{align} 

\subsection{Designing the master function}
\label{sec:fmaster}
In the following, we formulate the primitive variable recovery
as a root finding problem for a suitable master function.
To this end, we employ the following 
design goals

\begin{enumerate}
\item The function should be one-dimensional.
\item It should be continuous.
\item It should always have exactly one root, even for unphysical values of the conserved variables.
\item It should be well behaved in the Newtonian limit.
\item It should be well behaved for zero magnetic field.
\item There should be a known interval which contains the root and 
on which the function is defined.
\item The root-finding procedure should not require
derivatives of the EOS.
\end{enumerate}

For our scheme, we use $\Tmu$ defined in \Eref{eq:defmuandx} 
as the independent variable to solve for, 
i.e.~we will construct a function $f(\Tmu)$ which crosses zero where
$\Tmu$ is consistent with the conservative variables. The latter 
take on the role of fixed parameters. 
To construct $f$, we start with Eqs.~(\ref{eq:rbsqr}) and~(\ref{eq:qb})
and define
\begin{align}
\Fr^2 (\Tmu) 
  &= \Tr^2 x^2(\Tmu) + \Tmu x(\Tmu) \left( 1 + x(\Tmu) \right) \left( \Tr^l \Tb_l \right)^2 \label{eq:hatr}\\
\Fq (\Tmu) &= \Tq - \frac{1}{2} \Tb^2 - \frac{1}{2} \Tmu^2 x^2(\Tmu) \Tb^2 \Tro^2 \label{eq:hatq}
\end{align}
Next, we define functions for velocity and Lorentz factor
\begin{align}
\hat{v}(\Tmu) &= \min(\Tmu \bar{r}(\Tmu), v_0) \, , &
\hat{W}(\Tmu) &= \frac{1}{\sqrt{1 - \hat{v}^2(\Tmu)}} \label{eq:hatvel}
\end{align}
where $v_0$ is the upper velocity limit from \Eref{eq:vmax}.
Further, we define rest mass density and specific energy 
according to \Eref{eq:rhoraw} and \Eref{eq:epsraw}
\begin{align}
\hat{\rho} (\Tmu) &= \frac{\bar{D}}{\hat{W}(\Tmu)} \, , \label{eq:hatrho}\\
\hat{\epsilon}(\Tmu) 
&= \hat{W}(\Tmu) \left( \bar{q}(\Tmu) - \mu \bar{r}^2(\Tmu) \right) 
           + \hat{v}^2(\Tmu) \frac{\hat{W}(\Tmu)^2}{1 + \hat{W}(\Tmu)} \, , \label{eq:hateps}
\end{align}
provided that the results fall within the validity range of the EOS.
Otherwise, the density $\hat{\rho}$ is adjusted to the closest value
within the validity range for $\rho$, 
and $\hat{\epsilon}$ to the closest value
within the validity range for $\epsilon$ at adjusted density $\hat{\rho}$.
In \Eref{eq:hateps}, we express the term $W-1$ in a way that 
prevents large rounding errors in the case of small velocities.
Using the EOS, we compute the pressure, defining
\begin{align}
\begin{split}
\hat{P}(\Tmu) &= P(\hat{\rho}(\Tmu), \hat{\epsilon}(\Tmu)) \, , \\
\hat{a}(\Tmu) &= \frac{\hat{P}(\Tmu)}{\hat{\rho}(\Tmu)(1+\hat{\epsilon}(\Tmu))} \, , \\
\hat{h}(\Tmu) &= h(\hat{\rho}(\Tmu), \hat{\epsilon}(\Tmu)) \, .
\end{split}
\end{align}
To close the circle, we could now express $\Tmu$ itself as a function 
$\hat{\Tmu}(\Tmu)$ based on $\hat{W}(\Tmu)\hat{h}(\Tmu)$.
However, we find that this straightforward choice does not yield a
function respecting our design goals. One reason is that
under extreme conditions, the strong limitations introduced for 
$\hat{P}, \hat{\epsilon}, \hat{\rho}$, and $\hat{W}$ can cause 
severe kinks in the function. 

By trial and error, we find that using \Eref{eq:mualt} results in 
a very different master function $f(\Tmu)$ that is well suited 
to our purposes. It is given by
\begin{align}
f(\Tmu) &= \Tmu - \hat{\Tmu}(\Tmu) \label{eq:master}\\
\hat{\Tmu}(\Tmu) &= \frac{1}{\hat{\nu}(\Tmu) + \bar{r}^2(\Tmu) \Tmu} \label{eq:muhat}
\end{align}
As an additional minor modification, we compute the variable
$\nu \equiv h/W$ in two slightly different ways based on
Eqs.~(\ref{eq:defenthalpy}) and~(\ref{eq:epsraw}), defining
\begin{align}
\nu_A(\Tmu) &= \frac{\hat{h}(\Tmu)}{\hat{W}(\Tmu)}
= \left(1 + \hat{a}(\Tmu) \right) 
    \frac{1 + \hat{\epsilon}(\Tmu)}{\hat{W}(\Tmu)}  \label{eq:nuA}\\
\nu_B(\Tmu) 
&= \left(1 + \hat{a}(\Tmu)\right)\left(1 + \bar{q}(\Tmu) - \mu \bar{r}^2(\Tmu) \right) \label{eq:nuB}\\
\hat{\nu}(\Tmu) 
&= \max(\nu_A(\Tmu), \nu_B(\Tmu)) \label{eq:nuhat}
\end{align}
The motivation behind the second form $\nu_B$ is to reduce the kink introduced
by limiting $\hat{h}$, only limiting $\hat{a}$ instead, while allowing $\epsilon$
to change further. For this, $\left( 1 + \epsilon \right) / \hat{W}$ is computed 
directly from \Eref{eq:epsraw}, which does not involve the EOS.
The reason we use the smoother $\nu_B$ only when $\epsilon$ crosses the upper 
limit, but not when crossing the lower limit, is that decreasing $\hat{\nu}$ 
increases $\hat{\mu}$ and decreases the overall 
slope of the master function in some regimes, which is disadvantageous 
with respect to ensuring uniqueness.

\begin{figure}
  \begin{center}
    \includegraphics[width=0.95\columnwidth]{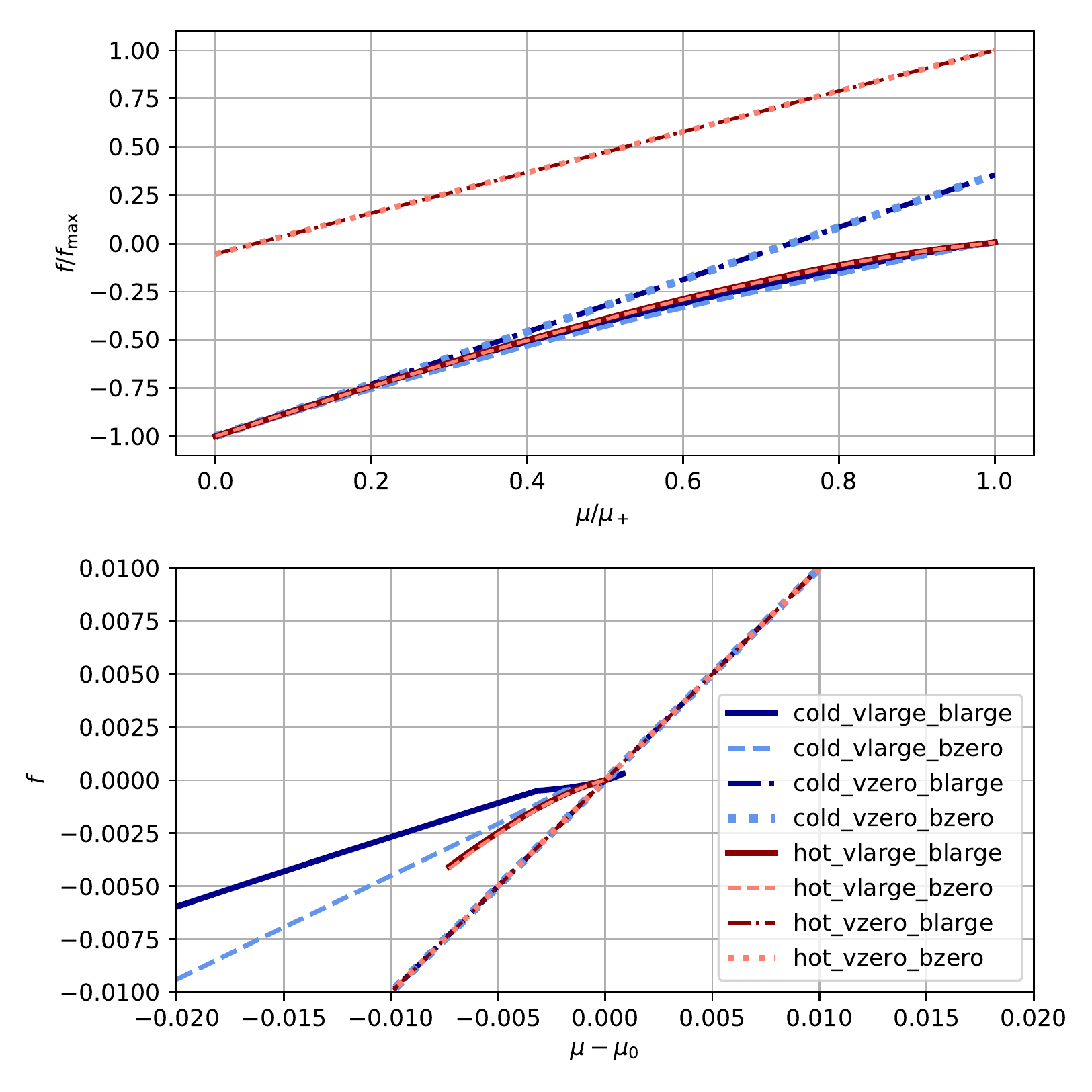}  
    \caption{Master function $f$ for different regimes, 
    combining: velocities $v=0$ and $v=0.99$ (labeled vlarge),
    magnetic field scale $b=0$ and $b=2$ (blarge),
    internal specific energy $\epsilon_{th} = 0$ (cold) and $10$ (hot), where
    $\epsilon_{th}$ denotes the difference to the zero temperature case.
    Velocities are oriented orthogonally to the magnetic field, 
    which is the most difficult case.
    In the top panel, the independent variable $\mu$ is scaled to the 
    initial root bracket $\mu_+$, and the function is scaled to the maximum value
    over the interval shown.
    The lower panel shows the behavior near the root $\tilde{\mu}_0$.  
 }
    \label{fig:master_func}
  \end{center}
\end{figure}

Examples for the master function resulting for different regimes are shown in 
\Fref{fig:master_func}. As one can see, the master function is almost linear
unless the Lorentz factor or magnetic scale $b$ are large, but it remains well behaved even then.
Note that we only show the root bracketing interval that is constructed 
in the next section. Beyond this interval, the function can have strong kinks.

\subsection{Existence of solution}\label{sec:exist}
In the following, we prove that the master function always has a root, not just for
valid evolved variables, but also for invalid ones. To this end, we construct an interval 
over which the master function changes sign. 
We start by defining an auxiliary function
\begin{align}
f_a(\mu) &= \mu \sqrt{h_0^2 + \Fr^2(\mu)} - 1  \label{eq:deffaux}
\end{align}
This is a smooth function which does not require evaluation of the EOS, using only the 
EOS-specific global lower enthalpy bound $h_0$.
It is easy to show that $f_a$ is strictly increasing and has exactly one root $\mu_+$ 
in the interval $(0,h_0^{-1}]$.
We will show that $\mu_+$ constitutes an upper bound for the root of the master function $f$.
For $f_a(\mu_+)=0$, we find
\begin{align}
\begin{split}
\mu_+ \Fr(\mu_+) 
&= \frac{\Fr(\mu_+)}{\sqrt{h_0^2 + \Fr^2(\mu_+)}} \\
&\le \frac{\Tr}{\sqrt{h_0^2 + \Tr^2}} = v_0  \, , 
\end{split}
\end{align}
where we used $r\ge \Fr$ and the monotonicity of the above expression 
with respect to $\Fr$. From \Eref{eq:hatvel}, we find 
$\hat{v} (\mu_+) = \mu_+ \Fr(\mu_+) \le v_0 < 1$.
Further, $f_a(\mu_+)=0$ implies that
\begin{align}
\mu_+ h_0 &= \sqrt{1 - \hat{v}^2(\mu_+)} = \hat{W}^{-1}(\mu_+)
\end{align}
Using the definition \Eref{eq:nuhat} of $\hat{\nu}$, which implies $\hat{\nu} \ge \nu_A$,
we can write
\begin{align}
\mu_+ \hat{\nu}(\mu_+) 
&\ge \mu_+ \frac{\hat{h}(\mu_+)}{\hat{W}(\mu_+)}
\ge \mu_+ \frac{h_0}{\hat{W}(\mu_+)} \label{eq:motg}\\
&= \hat{W}^{-2}(\mu_+) = 1 - \mu_+^2 \Fr^2(\mu_+) 
\end{align}
Hence,
\begin{align}
1 &\le \mu_+ \left( \hat{\nu}(\mu_+) + \Fr^2(\mu_+) \mu_+ \right) = \mu_+ / \hat{\mu}(\mu_+)
\end{align}
Using the definition of the master function, \Eref{eq:master}, we find that
$f(\mu_+) \ge 0$, and, trivially, $f(0) < 0$.
Since $f$ is continuous, it has at least one root in the interval $(0,\mu_+]$.
From \Eref{eq:motg}, we also find that the root is strictly below $\mu_+$
unless $\hat{h}(\mu_+) = h_0$.

Conveniently, the interval  provides a useful 
initial bracketing for the root finding algorithm. 
Although finding $\mu_+$ requires another numerical root solving, 
the computation of $f_a$ does not require the expensive evaluation of the EOS.
Moreover, determining $\mu_+$ is not required if $\Tr < h_0$. 
In this case, $\hat{v}(1/h_0) < 1$ and $f_a(1/h_0) > 0$, hence one can 
safely use $(0,1/h_0]$ to bracket the root.

The main reason to expend this effort is to ensure that $\hat{v} \le v_0 < 1$.
Beyond $\mu_+$, the cutoff in \Eref{eq:hatvel} can induce a strong kink
in the master function, reducing efficiency of the main root finding.
With the tight upper limit $\mu_+$, the only reason for the cutoff is to 
make absolutely certain that not even rounding errors in ultra-relativistic 
cases can lead to not-a-number results. Finally, being able to use 
$\hat{v} \le v_0$ simplifies the uniqueness proof in the next section.

\subsection{Uniqueness of solution}\label{sec:uniqe}

Uniqueness of physically valid solutions is obviously important for any evolution scheme
based on the conserved variables considered in this work. 
For the purpose of our recovery scheme, we require in addition that 
\begin{enumerate*}[label=(\roman*)]
\item for valid evolved variables, the master function has no 
additional roots corresponding to invalid solutions, and 
\item for invalid evolved variables, it still has exactly one root.
\end{enumerate*}
In this section, we will prove that the above conditions are met.

We first compute the derivative of the master function. 
Differentiation and straightforward algebraic computations yield
\begin{align}
\frac{\mathrm{d} x}{\mathrm{d} \mu} &= - x^2 b^2\\
\frac{\mathrm{d} \Fr}{\mathrm{d} \mu} &= - \frac{(1-x) x^2}{\mu \Fr} \Tro^2 \label{eq:drbardmu} \\
\frac{\mathrm{d} \Fq}{\mathrm{d} \mu} &= -(1-x)x^2 \Tro^2,  \label{eq:dqbardmu} \\
\frac{\mathrm{d} }{\mathrm{d}\mu} (\mu \Fr(\mu)) 
  &= \frac{1}{\Fr} \left( x^3 \Tro^2 + \Trp^2 \right) \ge 0
\end{align}
Since $\mu \Fr(\mu)$ is monotonically increasing and we have shown 
in \Sref{sec:exist} that $\mu_+ \Fr(\mu_+) \le v_0$, 
it follows that for $\mu\le \mu_+$, 
\Eref{eq:hatvel} reduces to $\hat{v}(\mu) = \mu \Fr(\mu)$.
From this, we find 
\begin{align}
\frac{\mathrm{d}}{\mathrm{d}\mu} \ln(\hat{W}) 
&= \hat{W}^2 \mu \left( x^3 \Tro^2 + \Trp^2 \right) \label{eq:dhatwdmu}
\end{align}
In the following, we assume that the density $\Fcrmd / \hat{W}$ does not 
leave the allowed range of the EOS. This corner case is discussed
in Appendix~\ref{app:deriv}.
We then obtain
\begin{align}
\frac{\mathrm{d} }{\mathrm{d}\mu} \ln(\hat{\rho}) 
 &= - \frac{\mathrm{d}}{\mathrm{d}\mu} \ln(\hat{W}) \label{eq:dhatrhodmu}
\end{align}
So far, we have computed derivatives of quantities that do not depend on \Eref{eq:hateps} and therefore we
did not need to consider the limiting of $\epsilon$ to the valid EOS range.
For the derivative of $\hat{\epsilon}$, we first consider the case in which $\epsilon$ 
computed by \Eref{eq:hateps} is in the valid range. We then find
\begin{align}
\frac{\mathrm{d}}{\mathrm{d}\mu} \hat{\epsilon} &=
\left( 1 + \hat{\epsilon} - \frac{1}{\hat{W}\mu} \right)
\frac{\mathrm{d}}{\mathrm{d}\mu} \ln(\hat{W}) 
\end{align}
At a solution, $\mu\hat{W}\hat{h} = 1$, and we obtain
\begin{align}
\frac{\mathrm{d}}{\mathrm{d}\mu} \hat{\epsilon} &=
-\frac{\hat{P}}{\hat{\rho}}
\frac{\mathrm{d}}{\mathrm{d}\mu} \ln(\hat{W}) 
= \frac{\hat{P}}{\hat{\rho}^2} \frac{\mathrm{d}}{\mathrm{d}\mu} \hat{\rho} \label{eq:hatadiab}
\end{align}
This means that changes of density $\hat{\rho}$ and specific energy $\hat{\epsilon}$ 
are adiabatic when varying $\mu$ near a solution, i.e. the derivative of specific 
entropy $\hat{s}$ is zero.
For the case in which $\epsilon$ computed by \Eref{eq:hateps} is below the valid range 
of the EOS, $\hat{\epsilon}$ is set to the lower bound of the validity range, 
$\epsilon_\text{min}(\hat{\rho})$,
which is the zero temperature limit according to our EOS requirements.
Consequently, $\hat{\rho}(\mu)$ and $\hat{\epsilon}(\mu)$ follow a curve of 
constant $s$.
In both of the above cases we can therefore compute the derivative of $\hat{h}$ 
using the adiabatic soundspeed given by \Eref{eq:csound} and the derivative of 
density given by \Eref{eq:dhatrhodmu}, obtaining
\begin{align}
\frac{\mathrm{d}}{\mathrm{d}\mu} \ln{\hat{h}} 
&= - \hat{c}_s^2 \frac{\mathrm{d}}{\mathrm{d}\mu} \ln{\hat{W}} 
\end{align}
where $\hat{c}_s = c_s(\hat{\rho},\hat{\epsilon})$.
In both cases, $\hat{\nu} = \hat{\nu}_A = \hat{h}/\hat{W}$, and we find
\begin{align}
\frac{\mathrm{d}}{\mathrm{d}\mu} \ln{\hat{\nu}} 
&= - \left(1 + \hat{c}_s^2 \right) 
\frac{\mathrm{d}}{\mathrm{d}\mu} \ln{\hat{W}} \, , \label{eq:hatnuderiv}
\end{align}
At a solution, the derivative of the master function becomes
\begin{align}
\begin{split}
\frac{\mathrm{d}}{\mathrm{d}\mu} f(\mu)
&= 1- \hat{v}^2 + \hat{v}^2 \left(1 - c_s^2 \right) \frac{x^3 \Tro^2 + \Trp^2}{\Fr^2} \\
&\ge 1 - \hat{v}^2 > 0 
\end{split} \, , \label{eq:masterderiv}
\end{align}
as shown in Appendix~\ref{app:deriv}.
The requirement $c_s<1$ is therefore sufficient to guarantee
uniqueness of the root for all velocities and magnetic fields.
Note that for superluminal soundspeeds, we cannot prove uniqueness
unless $v^2 c_s^2 < 1$. In that case, the proof of existence still holds, 
but is is not known if the physical solution becomes ambiguous.

So far, we have not addressed the corner case where the specific energy is above the valid 
range of the EOS. In that case, $\hat{\nu}$ is computed from \Eref{eq:nuB}.
A straightforward computation (see Appendix~\ref{app:deriv}) reveals that uniqueness 
is always ensured under the condition that
\begin{align}
\frac{A(\rho)}{1 + \hat{a}} \frac{\partial a}{\partial \epsilon} \le 1 - c_s^2 
\label{eq:req_epsmax}
\end{align}
The function $A$ is defined by the relation
\begin{align}
\frac{A(\rho)}{\rho} &=
\frac{\mathrm{d}}{\mathrm{d}\rho} \epsilon_\text{max}(\rho) 
- \frac{P}{\rho^2}  \label{eq:defadiabA}
\end{align}
It is related to the change of specific entropy along the upper validity 
range $\epsilon_\text{max}$ of the EOS. For $A=0$,
the equation above reduces to the thermodynamic condition for adiabatic change.
The condition given by \Eref{eq:req_epsmax} does not seem to be very restrictive. 
If in doubt, one can always use a boundary with $A=0$ (constant specific entropy) to 
guarantee uniqueness in all cases. In practice, we encountered no problems using
upper validity bounds defined by either constant temperature or constant $\epsilon$.

\subsection{Guaranteed accuracy} \label{sec:expacc}
Since the root of the master function is determined numerically, 
we require a criterion to stop the iteration once sufficient accuracy 
is reached. What is sufficient depends on the other errors present in a 
numerical evolution scheme. We will discuss evolution errors 
in \Sref{sec:evol_err}.
In this section, we discuss the error propagation of the root finding 
accuracy to quantify the accuracy of the recovered primitives.

However, we first need to specify how the final result is computed 
from the outcome of the last root finding iteration. This involves a design decision, 
since the available variables 
$\mu, \hat{\mu}, \hat{\nu}, \Fr, \Fq, \hat{v}, \hat{W}, \hat{\rho}, \hat{\epsilon}, \hat{h}, \hat{P}$
allow us to compute the primitives in many different ways, which lead to different
error propagation.
Here, we use 
$\hat{W}, \hat{\epsilon}, \hat{\rho}, \hat{P}$ directly, which turns out to be a good choice in terms 
of error propagation.
To reconstruct the velocity vector, we use the expression
\begin{align} 
\hat{v}^i &= \mu \Fr^i 
      = \mu x \left( r^i + \mu \left(b^lr_l\right) b^i \right) \, ,  \label{eq:velvecfinal}
\end{align}
which is just \Eref{eq:rfvec} rearranged to avoid degeneracy for the case $b=0$.
It is easy to verify that the Lorentz factor $W(\hat{v}^2)$
corresponding to \Eref{eq:velvecfinal} is exactly $W(\hat{v}^2)=\hat{W}$.
Since $\hat{\rho} = D / \hat{W}$, the conserved density $\hat{D}=\hat{\rho}W(\hat{v}^2)$ computed 
from the recovered primitive variables $\hat{\rho}, \hat{v}^i$ agrees exactly with the original one.

In the following, we only consider 
the case where the solution is in the validity region of the EOS.
For invalid solutions, the accuracy 
of the solution is less relevant since in this case the cause is the evolution error
and the result will either be corrected to the valid range or the simulation aborted.
The error introduced by such corrections will be discussed in \Sref{sec:random}.

Assuming the root of $f(\mu)$ was determined numerically to an accuracy of $\delta\mu$,
we now estimate the resulting accuracy of the primitive variables to linear order,
computing, e.g., ${\delta \hat{W} = \delta\mu\, \mathrm{d} \hat{W} / \mathrm{d}\mu}$.
We already evaluated the first derivatives at a root of $f$ in \Sref{sec:uniqe}.
From those, we obtain
\begin{align}
\frac{\delta\hat{W}}{\hat{W}} &\le \hat{v}^2 \Delta \, ,&
\Delta &\equiv \hat{W}^2 \frac{\delta\mu}{\mu} \label{eq:acc_w}\\
\frac{\delta\hat{z}}{\hat{z}} &\le \Delta \, , &
\frac{\delta\hat{v}}{\hat{v}} &\le \frac{|\delta\hat{v}^i|}{\hat{v}} \le  \frac{\Delta}{\hat{W}^2} \\
\frac{\delta\hat{\rho}}{\hat{\rho}} &\le \hat{v}^2 \Delta \, , &
\frac{\delta\hat{h}}{\hat{h}} &\le \hat{v}^2 \Delta\\
\frac{\delta\hat{\epsilon}}{1+\hat{\epsilon}} &\le \hat{a} \hat{v}^2 \Delta \, ,&
\frac{\delta\hat{\epsilon}}{\hat{\epsilon}} 
&\le \left(1+\hat{\epsilon}\right)\frac{\hat{a}}{\hat{\epsilon}}   \hat{v}^2 \Delta \label{eq:acc_eps}\\
\frac{\delta\hat{\rho_E}}{\hat{\rho_E}} &\le 2 \Delta \, ,&
\frac{\delta\hat{P}}{\hat{P}} 
&\le \hat{v}^2 \left(1 + \hat{a} \right) 
          \frac{\hat{c}_s^2}{\hat{a}} \Delta \, , \label{eq:acc_press}
\end{align}
where $|\delta\hat{v}^i|$ denotes the norm given by the 3-metric of the vector $\delta v^i$.
The error in the recovered primitive variables corresponds to errors $\delta q, \delta S_i$ of 
conserved variables.
We estimate these errors to linear order, by inserting $\hat{\rho},\hat{h},\hat{P},\hat{v}^i$ into 
equations~(\ref{eq:def_dens}--\ref{eq:def_cmom_tot}) and~(\ref{eq:idealmhd}) and then evaluate 
the first derivatives with respect to $\mu$ at the root. We obtain the following scaling
\begin{align}
\frac{|\delta S_i|}{|S_i|} &\le \Delta \, , &
\frac{\delta(1+q)}{1+q} &\le 4 v^2 \Delta \label{eq:acc_cons}
\end{align}
We find that the accuracy in $\mu$ required for a fixed relative error of the primitives 
increases with increasingly relativistic velocities.
On the other hand, the magnetic scale $b$ has no impact on the error bounds.
It is also worth noting that the error $\delta s$ of the
the specific entropy $s$ is zero (to linear order in $\delta\mu$) because
the variation of $\hat{\rho},\hat{\epsilon}$ with respect to $\mu$ is adiabatic (see \Sref{sec:uniqe}).
Finally, we note that the above error bounds do not include numerical rounding
errors. Those will be discussed in \Sref{sec:tests}.

%===========================================================================
\section{Enforcing Validity}
\label{sec:enforce}
In typical numerical simulations, the evolved magnetohydrodynamic 
variables frequently reach an invalid state at some points, mainly due to 
ordinary numerical error, but also external influences such as
gauge pathologies near the centers of newly formed black holes.  
Often, such 
violations are harmless and can be corrected. Any such correction turns unphysical
conditions into regular evolution errors, and obviously different prescriptions will lead 
to different errors, both in magnitude and in character. 
Although correcting violations should be regarded as part of the evolution scheme,
some basic point-wise corrections can be incorporated into the primitive recovery
code, granting it power to change the evolved variables.
The following effects cause typical harmless violations:
\begin{enumerate}
\item When evolving zero-temperature initial data, arbitrary small evolution errors
can lead to evolved variables that correspond to a fluid energy density below
the zero-temperature limit. 
\item At numerical grid points at the surface of neutron stars moving through vacuum, 
mass and energy densities during a single timestep can drop by orders of magnitude or even become 
negative. Although the absolute errors of the conserved variables remain small
compared to the global scales of the system, the resulting local error 
of the specific internal energy and velocity can become huge and lead to an
invalid state. The effect is alleviated over time because the errors tend to heat 
the outermost layer of NS surface, creating a hot atmosphere that reduces the 
density gradient.
\item During collapse to a black hole, mass density and/or temperature might leave
the range covered by the given EOS, arriving at a state that is not unphysical but
cannot be evolved further.
This typically occurs in regions already inside the 
horizon or about to be engulfed by a rapidly expanding apparent horizon.
\item
The coordinates near a black hole center are strongly stretched for gauges like
the puncture gauge, and the surroundings are extremely under-resolved numerically.
Under those conditions, all kinds of numerical instabilities can occur
for the combined magnetohydrodynamical and spacetime evolution system.
\end{enumerate}

\subsection{Simple corrections}
\label{sec:correct}

By design, our primitive variable recovery scheme is able to deal also
with invalid input. As a side-effect, we obtain a projection onto 
the valid regime, by simply recomputing the evolved variables from
the recovered primitives. As described in \Sref{sec:fmaster}, the scheme 
always yields a pair $\hat{\rho}, \hat{\epsilon}$ 
such that $\hat{\epsilon}$ is within the validity 
range of the EOS at $\hat{\rho}$. 

We first consider the important case in which the raw value of 
$\epsilon$ is below the valid range. In this case, only the recomputed
conserved energy $\tau$ changes, while $\Tcmom$ and $\Tcrmd$ stay the same.
This can be seen as follows. The only variable through which the 
adjustment of $\epsilon$ to the valid range impacts the master 
function is $\hat{\nu}$. For the case at hand, \Eref{eq:nuhat} implies 
$\hat{\nu}=\nu_A$. Furthermore, the conserved energy $\tau$ enters exclusively 
through \Eref{eq:hateps}. Therefore, if $\Tced$ is adjusted such that 
\Eref{eq:hateps} yields the range-limited value for $\hat{\epsilon}$, 
we arrive at the same primitive variables without adjustment. 

For the case in which the energy is above the validity range of the EOS,
all recomputed conserved variables can change. One could prevent this
by always using $\hat{\nu} = \nu_A$, but not without changing the behavior
of the master function away from the solution. However, this case is less 
important, because this correction should only be allowed at low-density 
fluid-vacuum boundaries (NS surfaces) or inside horizons.

In the interiors of black holes, it becomes necessary to employ a 
more lenient error policy than outside. 
Although physical effects cannot propagate out of the horizon, 
violations of the constraint equations and gauge effects impact the 
exterior. Therefore, one cannot allow any runaway instability inside
the horizon. For the matter part, this mainly concerns energy and 
momentum, since the total baryon number is conserved in finite
volume schemes (artificial atmosphere corrections aside), and the mass density remains finite.
The energy can be limited by allowing the aforementioned correction
to the EOS range inside horizons even at high densities. 

This leaves the momentum.
For pure hydrodynamic simulations, limiting the velocity proved effective
to prevent runaway instabilities near the BH center. This was employed for 
the simulations in \cite{Galeazzi:2013:64009}, by rescaling the velocity
to stay within a given limit. For MHD simulations, this approach has a
side-effect. Since the reconstructed electric field depends on the velocity
via \Eref{eq:idealmhd}, it will also change. That might be problematic or not,
depending on the evolution scheme. The evolution of the EM field
might be problematic in this regime in any case. However, addressing
such problems is clearly not inside the scope of the primitive variable 
recovery, since it operates point-wise and cannot change electric
or magnetic fields in any reasonable way.

Another correction often applied is to enforce a minimum mass density,
also called artificial atmosphere.
There are two motives. One is the wish to use a tabulated EOS that does not 
include zero density (this might be achieved more consistently by extending
the range to zero via analytic expressions). The more fundamental motive
is that the hydrodynamic evolution equations break down in vacuum.
In purely hydrodynamic simulations, it is common to set the 
atmosphere velocity to zero with respect to the simulation's coordinate 
system, in order to prevent an unphysical influx of matter. 
In ideal MHD simulations, the situation is more complicated because 
the electric field is tied to the velocity via \Eref{eq:idealmhd}.
Therefore, the atmosphere velocity prescription should be the domain of 
the evolution scheme and not of the primitive recovery.

%===========================================================================
\section{Performance}
\label{sec:perf}
In the following we assess how well our scheme performs in practice.
For this, we subject a reference implementation to a series of numerical
tests. These stand-alone tests simulate conditions relevant for the 
future use of the scheme in actual numerical simulation codes.

Our tests aim to validate that the following requirements are met. 
First, the scheme should not fail to converge for any input we expect to 
occur in numerical simulations of binary neutron star mergers (or supernova 
core collapse, which has similar demands). 
Second, within the physical regimes that may occur in the above scenarios, 
the errors caused by the primitive recovery should be insignificant compared 
to the ones caused by modern evolution schemes.

Lorentz factor $W$, magnetic scale $b$, and specific energy $\epsilon$ 
are the most important variables governing the behavior of the primitive recovery. 
To get an indication of what values to expect for $b$, we consider the merger 
simulations described in \cite{Ciolfi:2017:063016, Ciolfi:2019:023005} 
as a typical use case. The maximum
magnetic field strength reached after the merger by various amplification
mechanisms is ${\sim}10^{17}\usk\mathrm{G}$. The lowest density reached in 
the funnel along the axis after black hole formation is around 
${\sim}10^8\usk\gram\per\centi\meter\cubed$, although these conditions 
do not coincide.

The corresponding value $b \approx 100$ constitutes a reasonable 
scale up to which we demand robustness. 
A comparable value $b\approx 30$ corresponds to initial data with neutron 
stars possessing a magnetar-like exterior field strength $10^{15}\usk\mathrm{G}$ 
that extends into an artificial atmosphere with typical densities 
$10^5 \usk\gram\per\centi\meter\cubed$. 
In merger simulations, the regions with  magnetic fields amplified even further 
do not overlap with the lowest density regions. Combining a field
${\sim}10^{17}\usk\mathrm{G}$
with the aforementioned artificial atmosphere density, 
we find that $b\approx 10^3$ exceeds practical use cases by far
and demanding robustness up to this scale is not required.

Although the typical Lorentz factors we encounter in merger simulations 
are well below 10, one motivation for magnetized merger simulation is 
to observe the launch of a jet, and future applications might follow up such 
a jet to higher Lorentz factors. High Lorentz factors might also appear
inside black holes. Therefore, we demand robustness up to $W=10^3$.

Regarding the requirements for specific energy, we note that nuclear
matter at the highest densities possible in NSs can reach 
$\mathcal{O}(\epsilon) = 1$ at zero temperature. Although
temperatures can easily reach ${\sim}50\usk\mathrm{MeV}$ inside merging NS (see
\cite{Kastaun:2016} for example), 
this happens at high densities. The thermal contribution to $\epsilon$ 
stays well below unity (see, e.g., the appendix of \cite{Endrizzi:2018:043015}).
Furthermore, for states with $\epsilon\gg1$, the energy of the photon gas
dominates that of baryonic matter, and the evolution equations 
of magnetohydrodynamics are not applicable anymore.
It seems reasonable to demand accuracy and robustness of the primitive 
recovery up to $\epsilon\approx10$. 

In order to understand finite precision effects, we compare the accuracy that our reference 
implementation reaches to the expected accuracy derived in \Sref{sec:expacc}, using a 
value $\Delta=10^{-8}$. We assume that such an accuracy is negligible compared to the 
evolution error. Note, however, that in \Sref{sec:evol_err}, we show that the accuracy 
of $\epsilon$ (and hence $P$) related to the accuracy of the evolved variables 
deteriorates quickly for $b>1$, with a factor $\approx b^2 / \epsilon$. We therefore 
restrict tests of primitive recovery accuracy to the regime $b<5$ and caution
that one cannot trust the evolution scheme at higher values.

\subsection{Code Design}
\begin{figure}
\includegraphics[width=\columnwidth]{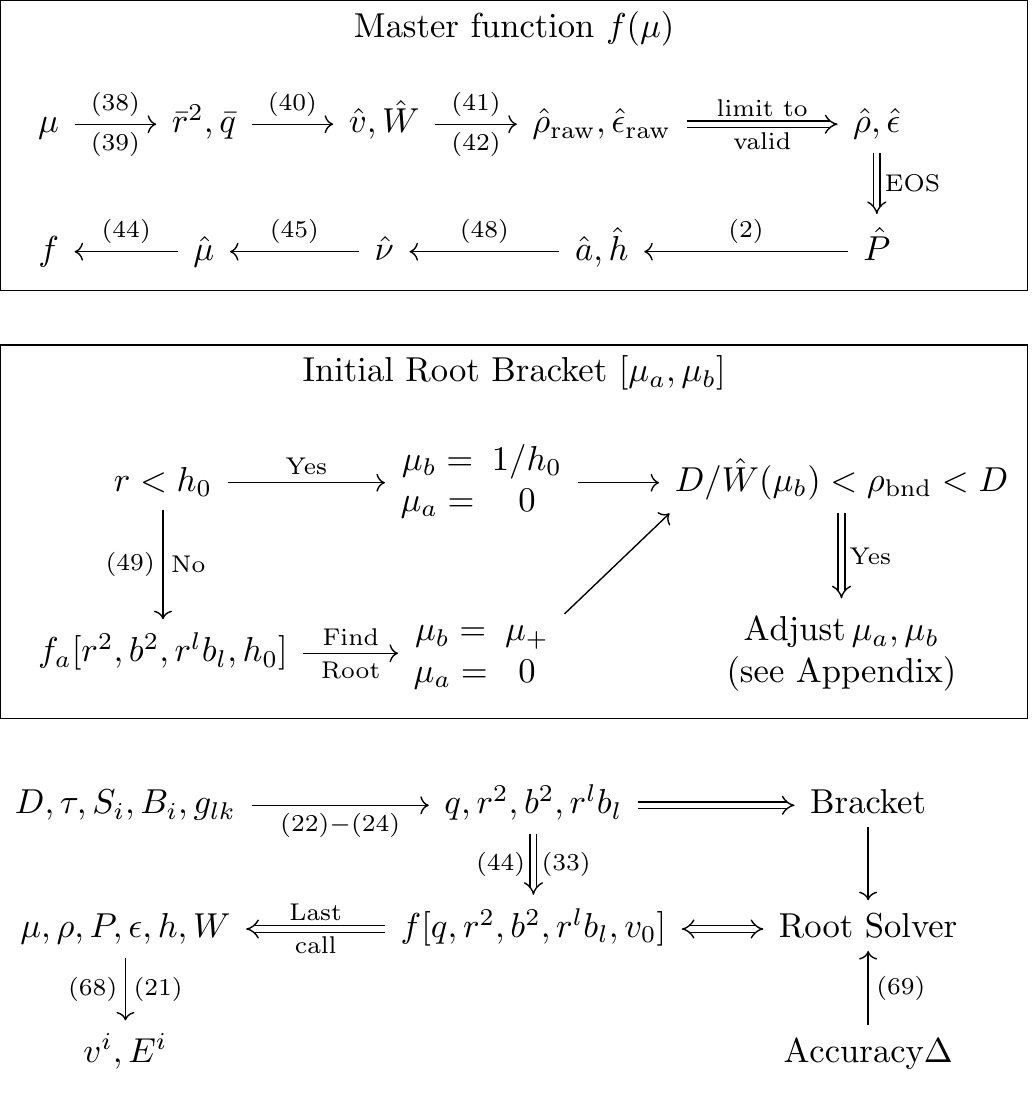}
\caption{Schematic overview of the recovery algorithm 
information flow and list of required equations. Arrows denote 
dependencies, including variables further up in the chain.
Double arrows denote computations that require information about 
the EOS validity bounds, the one double arrow labeled ``EOS'' refers to
evaluation of the EOS. Intermediate results
obtained during evaluation of the master function in the last root 
finding iteration are denoted ``Last call''. Square brackets refer 
to the list of fixed parameters (independent of $\mu$) during root solving. 
The density $\rho_\text{bnd}$ stands for any one of the upper and lower 
EOS validity bounds.
}\label{fig:flowchart}
\end{figure}

We created a reference implementation for the algorithm described
in \Sref{sec:formulation}. A schematic summary of the algorithm
is given in \Fref{fig:flowchart}.
We provide our reference implementation in the form of a C++ library 
named \texttt{RePrimAnd}. 
The library is not tied to any particular evolution framework,
allowing the use in arbitrary evolution codes.
It also contains a framework providing 
access to different types of EOS through a generic interface, ensuring
that the user code (such as our primitive recovery code) is completely 
EOS-agnostic. The generic interface 
also provides the EOS validity ranges and rigidly enforces our EOS 
requirements (see \Sref{sec:eos}).
The reference implementation is publicly available \cite{RePrimAnd}.

Our library provides different EOS types (not yet including fully tabulated
ones) including a hybrid EOS based on a tabulated cold part. 
Although the MS1 variant from \cite{Read:2009:124032} used for our test is given
analytically in the form of piecewise polytropic expressions, we evaluate it in
our tests as a tabulated cold part, in order to test the general
purpose code intended for production runs. We set the allowed range
of the EOS to 
$\rho_\mathrm{max} = 3\times 10^{15}\usk\gram\per\centi\meter\cubed, \epsilon_\mathrm{max} = 51$.

In order to find the root of the master function, our implementation uses
the \texttt{TOM748} algorithm \cite{TOMS748} provided by the \texttt{BOOST} library.
This root finding scheme is similar to the well known Brent-Dekker schemes,
but it uses inverse cubic instead of quadratic interpolation whenever possible,
improving convergence speed near the solution.  
It keeps the solution enclosed in a bracket, with extent converging 
in a limited amount of steps, and it does not make use of function derivatives.
The motivation for avoiding derivatives is that, in practice, tabulated
EOS tend to have very inaccurate partial derivatives, which is problematic
when using a derivative-based root solver such as Newton-Raphson. Our 
implementation therefore does not make use of the soundspeed or other 
derivatives. The root is determined to an accuracy specified by prescribing
$\Delta$ defined in \Eref{eq:acc_w}, where the error $\delta\mu$ is taken
as the size of the current tightest bracket of the root.

In the case where $\Tr \ge h_0$, we need to determine $\mu_+$ from the root of the auxiliary 
function $f_a$, for which we employ a standard Newton-Raphson root solver. 
This is unproblematic since $f_a$ is a smooth, monotonic, analytic function. 
We determine $\mu_+$ to an accuracy 
close to machine precision, and then increase $\mu_+$ by a multiple of the 
root solving accuracy to ensure the root of the master function is really 
contained.  

\subsection{Robustness and Accuracy}
\label{sec:tests}
Our main test validates both robustness and expected accuracy, by sampling
the primitive variable parameter 
space given by density, temperature/specific energy, magnetic scale $b$, 
and velocity.
We sample $z=Wv$ between $0$ and $1000$, magnetic scale $b$ from 
$0$ to $5$, and the specific thermal energy from 
$\epsilon_\mathrm{th}=10^{-4}$ up to $50$.
For the MS1 EOS, we sample the mass density from $10^{6}$ to 
$10^{15} \usk\gram\per\centi\meter\cubed$. For the ideal gas EOS, the 
mass density is irrelevant due to the scaling behavior of the EOS.
We use two orientations of the velocity, parallel and 
orthogonal to the magnetic field. The tests are performed both for the 
ideal gas and for the hybrid EOS, described in \Sref{sec:eos},
and we demand an accuracy $\Delta=10^{-8}$.

We verify that the algorithm always, without any exception, 
succeeds in recovering the correct solution for the valid input described above. 
Moreover, we 
create test cases to assure that input corresponding to energy 
outside the range possible for a given EOS is correctly classified 
as such.

To assess the accuracy, we compute the conserved variables from the primitives,
apply the primitive recovery algorithm, and compare the result to the 
original primitives. 
Further, we compute the conservatives from the recovered primitives
and compare them to the original conservatives. 
Our testsuite compares the observed accuracy for each individual primitive 
variable to the one expected
from \Eref{eq:acc_w} to \Eref{eq:acc_press}, and also the accuracy of the 
corresponding conserved variables to \Eref{eq:acc_cons}. 
When demanding an accuracy better than $\Delta \lesssim 10^{-7}$,
those bounds are exceeded either for high Lorentz factors or very small 
$\epsilon$ and $v$. We attribute the excess error to various rounding errors. 

We identify the most important rounding errors as follows. 
First, the master function is the difference of two values which can each be expressed 
only to machine precision. To get the impact on the root, we have to divide by the 
derivative of the master function, which in this case satisfies 
$f' \le 1 - v^2 c_s^2$. At the same time, we demand an accuracy $\Delta / W^2$.
For the highly relativistic case $W=10^3$ and around 16 digits machine precision,
this limits $\Delta>10^{-10}$. If the soundspeed approaches unity, the accuracy
is further limited.
Second, $\epsilon$ is computed by subtracting kinetic and magnetic energy density
from the total one. If $\epsilon$ is small compared to these, the cancellation error
causes a loss of significant digits. Analyzing \Eref{eq:hateps}, we find additional 
rounding errors of magnitudes $z^2/\epsilon$ and $b^2 W / \epsilon$ worse
than machine precision.

\begin{figure}
  \begin{center}
    \includegraphics[width=0.95\columnwidth]{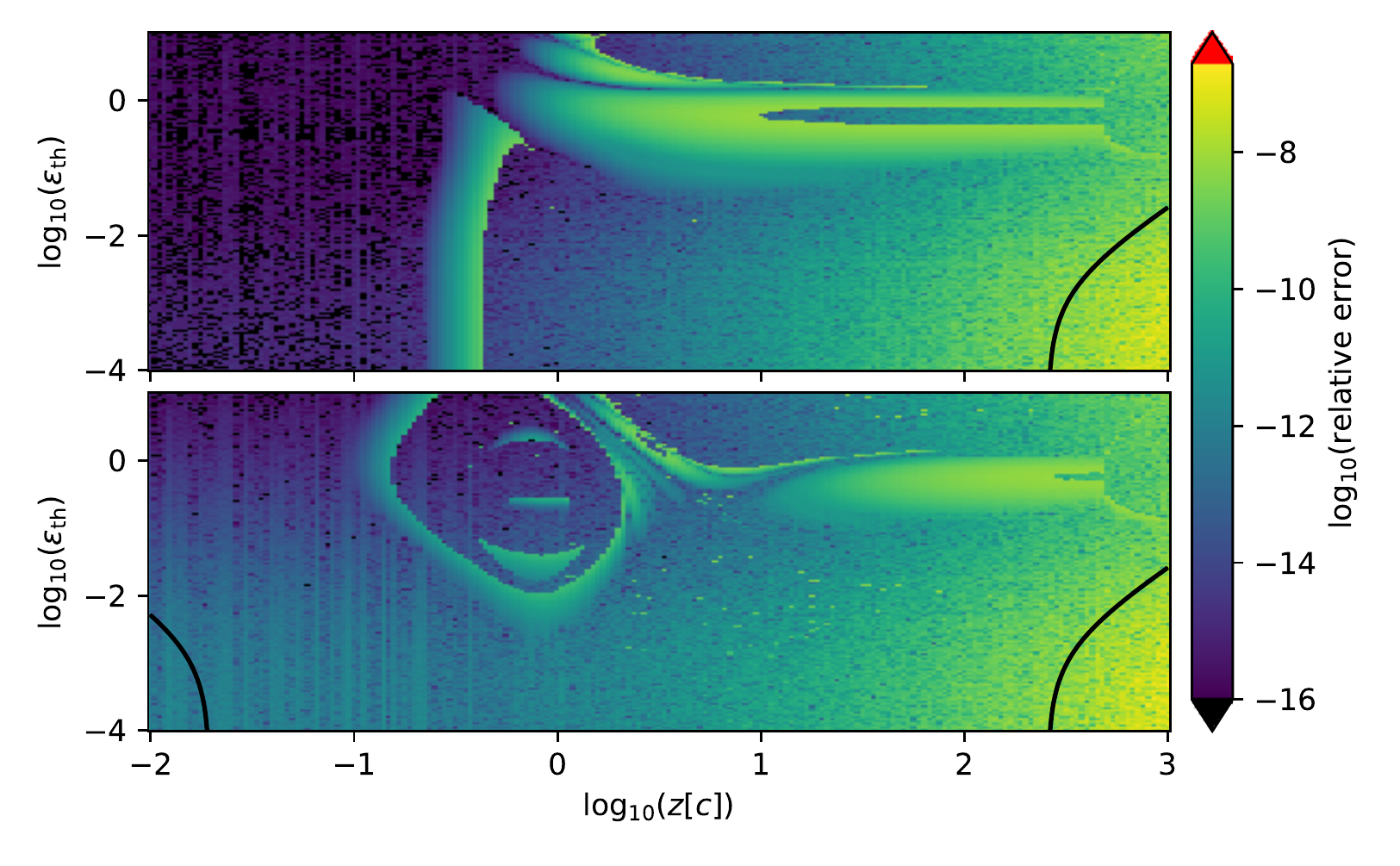}  
    \caption{Relative error of reconstructed pressure,
    as a function of specific thermal energy and velocity (the latter in terms
    of $z=Wv$). 
    The results were obtained for the case of the hybrid EOS (see text),
    at fixed mass density $6 \times 10^{12} \usk\gram\per\centi\meter\cubed$,
    demanding an accuracy $\Delta=10^{-8}$.
    The upper panel shows results for zero magnetic field, and the lower panel
    for the magnetically dominated case $b=10$.
    The solid lines mark the regions where expected errors related to 
    rounding start to exceed those related to root solving. 
    }
    \label{fig:acc_press_z_eps}
  \end{center}
\end{figure}

Taking into account both the regular errors predicted by  \Eref{eq:acc_w} 
to \Eref{eq:acc_press} as well as the main rounding errors discussed above,
the recovered accuracy is quantitatively within the expected bounds over the 
whole range of our test cases. \Fref{fig:acc_press_z_eps} shows the recovered 
accuracy for the pressure as well as the boundary where the errors caused by 
rounding start exceeding those caused by root finding. 

We do not expect rounding errors to be of practical importance. 
The rounding errors at low $\epsilon, v$
are very small and only dominate because the regular errors approach zero.
The rounding errors in the ultra-relativistic/highly magnetized regime 
are still not prohibitive, but will be dominated by the errors of the time 
evolution, which will be discussed in \Sref{sec:evol_err}.

\subsection{Efficiency}
\label{sec:speed}
In the following, we discuss the efficiency of our scheme on the level of the algorithm, 
while reserving benchmarks of the execution speed of the implementation within 
actual GRMHD simulations for future work.
We measure the computational efficiency of our algorithm in terms of calls to the EOS.
The motivation is that for a tabulated EOS including thermal and composition degrees
of freedom, a single EOS call is likely more expensive than the evaluation of the 
analytic expressions within our recovery scheme. The worst scenario is when the EOS 
is tabulated 
with temperature as one independent variable. Each EOS call then requires an inversion 
step to convert from $\epsilon$ to $T$. 

\begin{figure}
  \begin{center}
    \includegraphics[width=0.95\columnwidth]{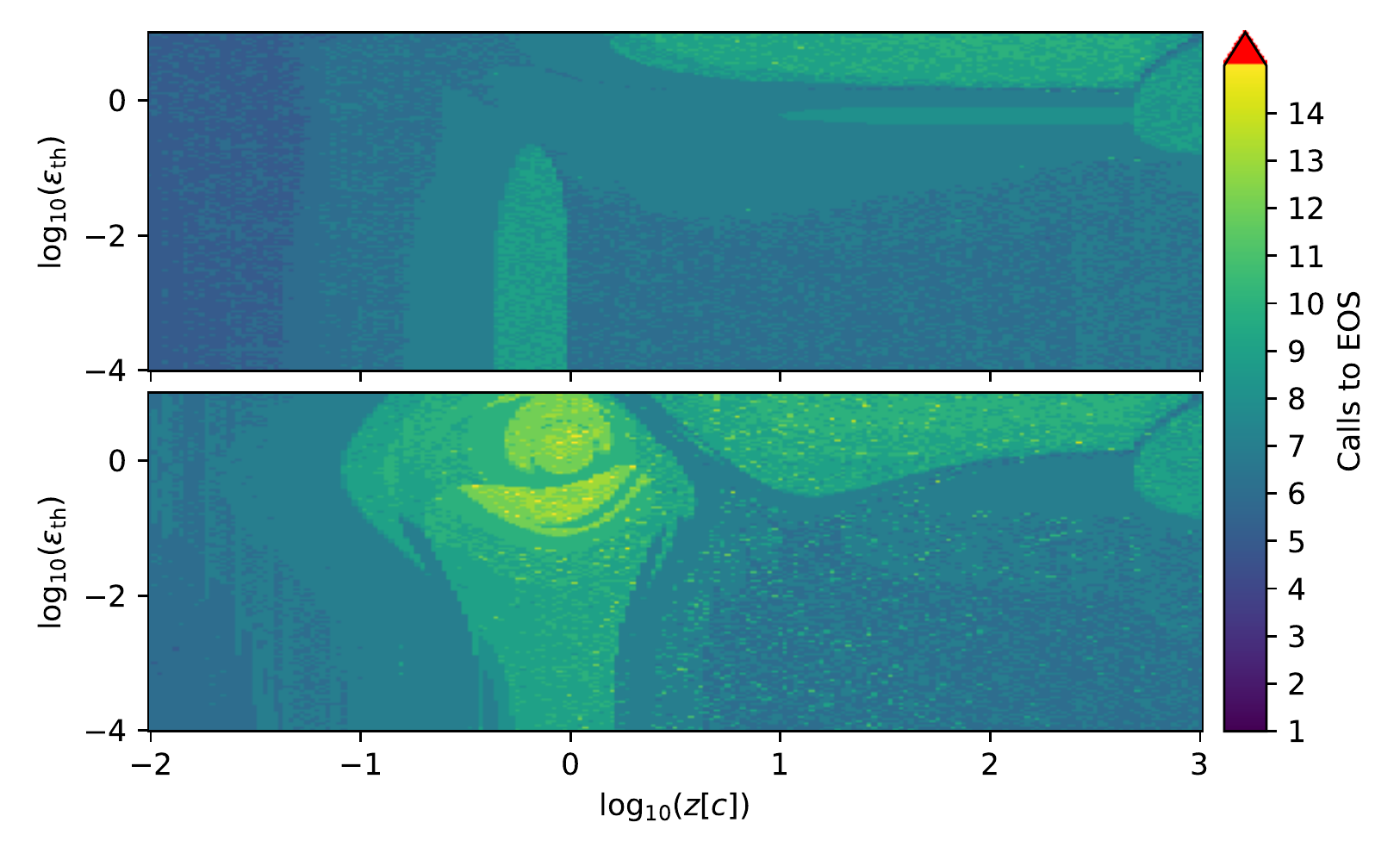}  
    \caption{Number of calls to the EOS required to reconstruct the primitives
    to accuracy $\Delta=10^{-8}$,
    as a function of specific thermal energy and velocity (in terms of $z=Wv$). 
    The results were obtained for the case of the hybrid EOS (see text) at
    a mass density $\rho=6 \times 10^{12} \, \gram\per\centi\meter\cubed$.
    The upper panel shows results for the magnetic scale $b=0$, and the lower panel
    for the magnetically dominated case $b=10$.
    }
    \label{fig:perf_eoshyb_z_eps}
  \end{center}
\end{figure}

\Fref{fig:perf_eoshyb_z_eps} shows how the efficiency varies with specific energy
and velocity, either for zero magnetic field, or with magnetic scale fixed to a large 
value of $b=10$.
We find that the efficiency does not degrade even for Lorentz factors up to 1000 
and magnetic scales up to $b=10$. At the density shown, $b=10$ corresponds to extremely
high magnetizations of orders $10^4$ (for $W=1$) to $10^7$ (for $W=1000$).

When considering the whole parameter space used in the unit tests (not just the cuts 
shown in the plots) and both EOS types, we find a maximum number of 23 calls to the 
EOS required to achieve an accuracy $\Delta=10^{-8}$. 
The maximum occurs for the ideal gas and only when both $\epsilon > 40$ and $b>2$, i.e., thermal energies 
much larger and magnetic energies larger than the rest mass density.

\begin{figure}
  \begin{center}
    \includegraphics[width=0.95\columnwidth]{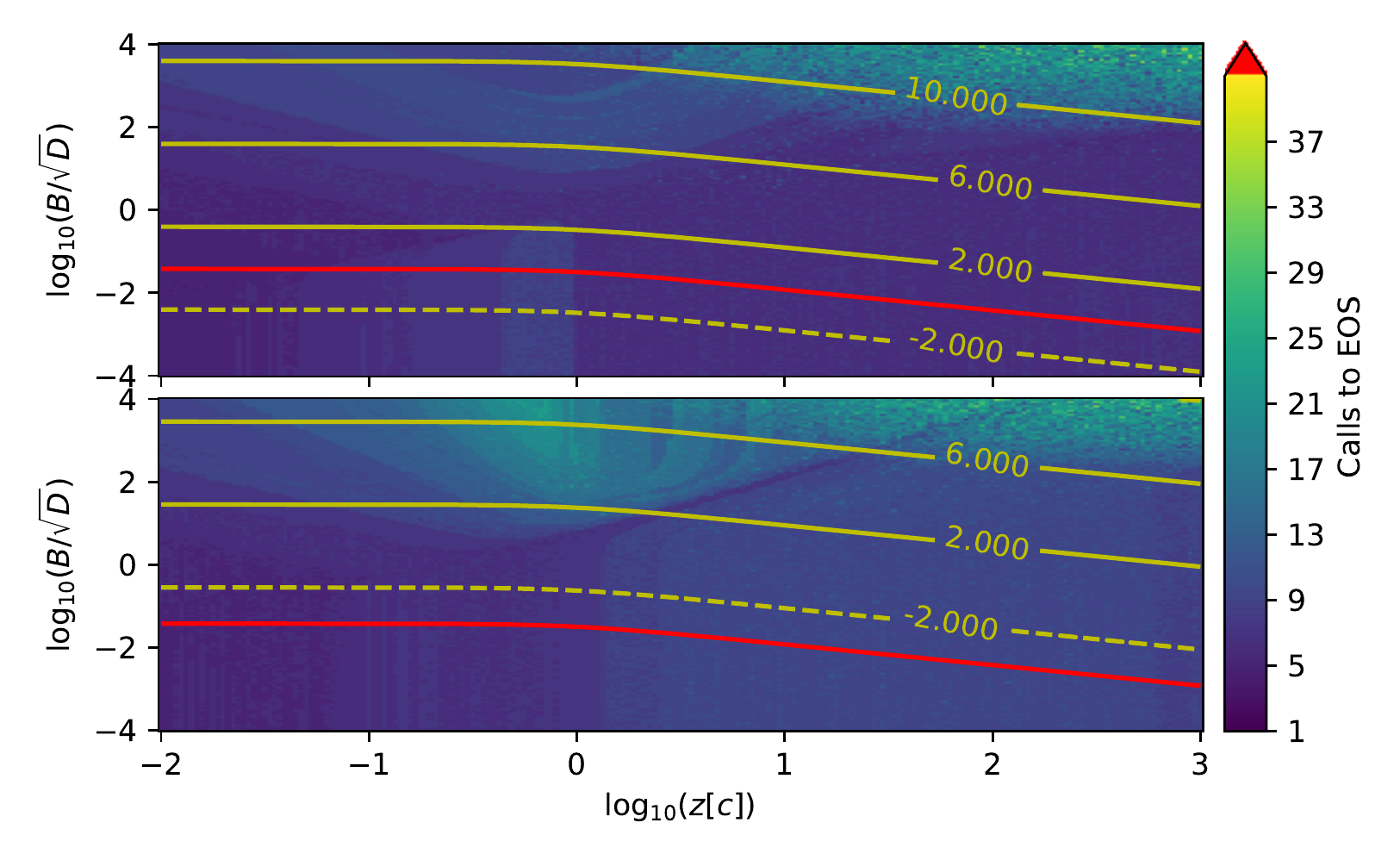}
    \caption{Number of calls to the EOS required to reconstruct the primitives
    to accuracy $\Delta=10^{-8}$,
    as a function of magnetic scale $b$ and velocity (in terms of $z=Wv$).
    The results were obtained for the case of the hybrid EOS (see text) at
    a mass density $\rho=6 \times 10^{12} \, \gram\per\centi\meter\cubed$.
    The upper panel shows results for cold matter $\epsilon_\text{th}=10^{-4}$, 
    and the lower panel for very hot matter $\epsilon_\text{th}=10$.
    For comparison, we also show the magnetization as contour lines of
    $\log_{10}(B^2/P)$. The red line marks a magnetic field strength
    $B=10^{16}\usk\mathrm{G}$.
    }
    \label{fig:perf_eoshyb_z_b}
  \end{center}
\end{figure}

In \Fref{fig:perf_eoshyb_z_b}, we show the efficiency with respect to 
velocity and magnetic scale $b$, taking the latter up to extreme values $b=10^4$,
far beyond any reasonable use case. We find that beyond $b=10$, 
the efficiency gradually starts to decrease. At $b=10^4, W=10^3$, we require around 40 
steps for $\Delta=10^{-8}$ (which implies  $\delta \mu / \mu \approx 10^{-14}$). 
At this point, the root solving convergence speed has decreased roughly to that of 
bisection. Still, we encounter no failures to converge even in this range.

Note that the extremes reached in our tests are rather pathological scenarios 
which are rarely encountered in simulations and are therefore not relevant
for numerical costs of simulations. In practice, 
we expect an average number of required calls below 10.

\subsection{Comparison with other schemes}
\label{sec:compare}

In the following, we compare our scheme to existing ones. 
We refer to \cite{Siegel:2018:71} for a comprehensive review
and numerical tests of previous schemes.
The main characteristics are listed in \Tref{tab:schemes}.

One important difference is the number of independent 
variables. Most of the existing schemes need to solve an
equation in two or three unknowns. This is a severe drawback.
First, it is difficult to ensure that the solution is found.
The Newton-Raphson (NR) schemes might not converge. Second,
robust but fast schemes that guarantee finding the solution
in a limited number of steps only exist for one-dimensional
root finding. Third, the recovery schemes based on NR require
an initial guess, which is typically taken from the previous
timestep during numerical evolution. This makes the methods more
unpredictable and more difficult to test, as they do not depend
on the conserved variables in a deterministic way.
As two of the existing schemes, our scheme is using one-dimensional 
root finding. Further, it also makes use of a tight initial 
bracketing interval proven to contain exactly one solution.

As demonstrated in \cite{Siegel:2018:71} 
all of the existing schemes can fail for Lorentz factors 10--1000,
depending on the magnetization.
Fig.~3 of \cite{Siegel:2018:71} shows the number of iterations or failure
to converge as function of magnetization and Lorentz factor, at fixed density
$10^{11}\usk\gram\per\centi\meter\cubed$ and $T=5\usk\mathrm{MeV}$ (thus  
$\epsilon < 1$).
For our scheme, $b$ is the most relevant measure for the magnetic field (but 
not necessarily for the other schemes). For 
comparison, the magnetization $10^{10}$ covered in the figure corresponds 
to values up to $b\approx 10^4$. As argued before, such values are  
outside the parameter space relevant for use in merger simulations.
Therefore, the failures at low velocity, but with magnetization around $10^9$ 
shown in Fig.~3 of \cite{Siegel:2018:71} are not problematic
in practice. The fact that our scheme showed no failure at $b = 10^4$ for 
the test shown in \Fref{fig:perf_eoshyb_z_b} is however reassuring regarding 
the numerical robustness.
The failures at relativistic velocities for lower magnetization  
shown in Fig.~3 of \cite{Siegel:2018:71} should however be regarded as problematic.
For our algorithm, the existence and uniqueness of the solution are 
proven analytically, and 
we successfully test our numerical implementation up Lorentz factors $W=1000$
in the whole parameter space described in \Sref{sec:tests}.

Strong magnetization is important for studying the engine of short gamma-ray bursts (SGRBs),
where ideally a very low density matter is subject to very strong 
magnetic fields. This regime is also problematic for the numerical time 
evolution itself. The ability of our scheme to distinguish reliably 
between valid and invalid evolved variables is therefore an important
advantage.

Fig.~2 of \cite{Siegel:2018:71} shows the average of the relative 
errors (called $\bar{\sigma}$ in the following) for 
$\rho, v^i, \epsilon$ as a function of temperature and density, for fixed
low magnetization $p_\mathrm{mag}/p = 10^{-3}$ and Lorentz factor $W=2$.
We note that for the nuclear physics EOS, one can ignore the top-left part 
of the plots in Fig.~2, because this low-density, high-temperature regime corresponds 
essentially to a photon gas not relevant for practical use. Using the energy 
density $\rho_\gamma\propto T^4$ of a photon gas, we find that the bound $\epsilon<10$ 
used in our tests corresponds to temperatures below a straight 
line (in the log-log plot) through 
$(10^{12} \usk\gram\per\centi\meter\cubed, 10^{12}\usk\kelvin)$
and $(10^4\usk\gram\per\centi\meter\cubed, 10^{10}\usk\kelvin)$.
For the ideal gas EOS, the temperature is instead defined 
assuming only baryons, such that $\epsilon<1$ throughout the figure.

In most of the physically relevant region, the recovery accuracy shown in 
\cite{Siegel:2018:71} seems more than sufficient for use in evolution schemes.
However, some existing schemes exhibit isolated regions where the accuracy 
degrades inexplicably. This is worrisome since the figure shows only a  
mildly relativistic twodimensional cut in the parameter space, and there 
is no guarantee that the accuracy will not deteriorate intolerably elsewhere.
In contrast, our scheme has the advantage of a theoretical model for the errors of 
each of the primitive variables, including the dominant finite precision errors.
This model was validated for our numerical implementation over the full 
parameter space $(\rho,W,\epsilon,b)$ as described in \Sref{sec:tests}.

Regarding the efficiency, the different tolerance measures allow only 
a rough comparison, using the number of root finding iterations
shown in Figs.~1 and~3 of \cite{Siegel:2018:71}. 
Note that the top-left part of the plots in Fig.~1 
is not practically relevant for the nuclear physics EOS,
as discussed above. As mentioned before, the problems at the highest 
magnetizations $>10^9$ in Fig.~3 of \cite{Siegel:2018:71} can be 
safely ignored for practical use. In case no failure occurs,
only the Newman scheme appears to be consistently requiring less than 
around 10 steps also for relativistic velocities. The others need 30 
iterations or more in certain regions of parameter space. 

Our scheme is guaranteed to converge in a finite 
number of steps because
the root finding algorithm performs bisection steps if needed. 
Our tests have shown that the efficiency does not degrade for large 
Lorentz factors ($W<1000$) or strong
magnetization ($b<100$), in contrast to most other schemes. The worst case 
scenario for our scheme seems to be extreme values of $\epsilon$. Even for 
essentially photonic states ($\epsilon=50$), it does not require more 
than 23 EOS calls in the regime $b<5,W<1000$.

Up to this point, we considered the convergence criteria used
during the iteration as an integral part of the different algorithms.  
Now we have to discuss if the different measures of success could bias
the comparison.
In \cite{Siegel:2018:71}, recovery was called successful when an average
error $\bar{\sigma} < 5\times 10^{-8}$ could 
be achieved. Comparing this to the root solving errors given by
Eqs.(\ref{eq:acc_w}) to~(\ref{eq:acc_press}), we find that 
$\bar{\sigma} < K \Delta$, where $K$ is a constant of order unity
\footnote{This ceases to be true near zero-crossings of $\epsilon$.  
Whether crossings can occur depends on the choice of the arbitrary 
constant $m_\mathrm{B}$ (see \Sref{sec:primvars}). This ambiguity is a general 
problem with the definition of $\bar{\sigma}$, which contains a term 
$\delta\epsilon / \epsilon$. It is not relevant for the following discussion, 
but might be a pitfall in the low-density, low-temperature
regime.}.
However, the term $\delta\epsilon / \epsilon$ is also very sensitive to 
unavoidable cancellation errors, as already discussed in \Sref{sec:tests}.
Taking into account the finite floating point precision of $q,r,b$, we find that
catastrophic cancellation decreases the number of valid digits for $\epsilon$ 
by around $\log_{10}(z^2/\epsilon)$ for the case $b=0, z^2 \gg \epsilon$, and 
by $\log_{10}(b^2/\epsilon)$ for $v=0, b^2 \gg \epsilon$.
For the ideal gas case shown in the upper panel of Fig.~3 in 
\cite{Siegel:2018:71}, $\epsilon$ and $P$ are constant and one can easily 
estimate the loss of precision. Assuming a typical machine precision
of 15 digits for $q,r,b$, rounding errors will make it impossible to 
reach the tolerance $\bar{\sigma}$ required in \cite{Siegel:2018:71}
at $W \gtrsim 2 \times 10^3$ (for small $B$) or at 
$P/P_\mathrm{mag} \gtrsim 5\times 10^8$ (for small $v$).
The above errors are a fundamental consequence of evolving 
the conserved variables, independent of the recovery scheme. 
Some, if not all of the ``failures'' near the upper boundary
of the ideal gas panels are therefore inevitable,
also for our scheme.

We test if our implementation can reach the $\bar{\sigma}$-tolerance 
from \cite{Siegel:2018:71} within our standard test domain introduced in 
\Sref{sec:tests}. 
In general, at $b=0$ our theoretical order of magnitude estimate 
predicts a larger impact of rounding errors for $\epsilon \lesssim 0.005 \cdot (W/10^3)^2$.
For the ideal gas, the tolerance $\bar{\sigma}$ was indeed only reached when
restricting $\epsilon$ above this estimate. Apart from this regime, the tolerance was 
reached in the whole standard test domain ($W \leq 10^3, b<5$) using $\Delta=10^{-8}$.
The same holds for the hybrid EOS.
We conclude that the impact of the cancellation error on the error 
measure $\bar{\sigma}$ is close to the theoretical minimum for our implementation. 
In contrast, all schemes shown in \cite{Siegel:2018:71} exceed 
the tolerance already below $W<10^3$ at small $B$, where our scheme succeeds. 

The above discussion indicates that $\bar{\sigma}$ might not be a good choice as a 
criterion for recovery failures. 
We also point out that the relative
error of $\epsilon$ is not a good measure for the error of the temperature, which is 
more closely related to the thermal part $\epsilon_\mathrm{th}$.
For practical applications, it seems preferable to
allow the inevitable accuracy loss for $\epsilon$ discussed above, recovering each primitive 
variable as accurately as theoretically possible. Whether or not the unavoidable
cancellation errors in $\epsilon$ are tolerable surely depends on the application and should be 
regarded as part of the post-recovery error policy.

As is pointed out in \cite{Siegel:2018:71}, comparing the number of EOS 
calls between different schemes does not directly translate
to numerical costs. The reason is that the schemes differ with respect
to the required form of the EOS.
Our scheme is using $P=P(\rho,\epsilon)$, while nuclear physics EOS tables
are typically given in terms of $T$ instead of $\epsilon$.  
It was argued in \cite{Siegel:2018:71} that it is advantageous if the 
master function directly uses the temperature as the independent variable, 
because otherwise each call to the EOS requires another root finding 
to determine $T$. We regard this as a shortcoming of the EOS 
implementation and advocate against basing the design of the primitive
recovery on internals of specific EOS implementations. A more natural solution
is to first create new tables in terms of $\epsilon$ (or a suitable analytic 
function thereof), by interpolating available nuclear physics tables. 
This allows one to choose the 
most robust recovery procedure without sacrificing speed. When using a lookup
table based on temperature, however, the scheme proposed in \cite{Siegel:2018:71} might 
indeed be faster than ours.
We point out, however, that the large speedup discussed in \cite{Siegel:2018:71} seems
to be based on a particularly wasteful implementation of the inversion 
$T(\epsilon)$ that can require up to hundred steps. In \cite{Galeazzi:2013:64009}, 
we used a discrete bisection in index space followed by inverse interpolation,
which requires $<10$ steps for realistic table sizes.

In contrast to \cite{Siegel:2018:71}, we do not test the scheme with a 
fully tabulated EOS and can therefore make no conclusive claims on robustness
and accuracy of the implementation in this case. 
However, the algorithm itself is guaranteed to find a solution.
We recall that the proof of existence does not rely on EOS properties except
for a lower bound on $h$. A table in conjunction with an interpolation 
method represents a well defined EOS. As long as this EOS respects the 
physical constraints listed in \Sref{sec:eos}, the 
uniqueness is also guaranteed. A careless 
implementation of a tabulated EOS that violates those constraints might 
however cause our algorithm to find wrong, unphysical solutions. For such
faulty EOS there might even exist several physical solutions. Furthermore,
our accuracy bounds are not guaranteed anymore because they rely on
the constraints as well. Still, we do 
expect the numerical root finding to converge to a solution. Even jumps in 
the master function would only reduce the efficiency of the 
TOM748 solver, possibly down to that of bisection. We recall that this root 
finding method does not use derivatives, and would therefore not be affected 
by an EOS implementation returning inaccurate numerical derivatives of the 
pressure. Such inconsistencies might affect those schemes 
in \cite{Siegel:2018:71} based on Newton-Raphson root solvers. However, the 
failures to converge exposed in \cite{Siegel:2018:71} also appear for the 
purely analytic ideal gas EOS, and can therefore not be attributed to possible 
faults in EOS tables.

\begin{table}
\caption{Main characteristics of different recovery schemes.
We list the independent variables used in the root finding (translated 
to our notation), the variables for which the EOS needs to be evaluated, 
whether the scheme requires derivatives of the EOS, 
whether the formulation allows a bound on the number of
iterations needed for finding the solution, and whether the scheme requires
to provide an initial guess for the solution.}
\label{tab:schemes}
\begin{tabular}{c|c|c|c|c|c}
\hline 
Scheme & Independent & EOS  & EOS   & Steps & Guess \\
       & variables   & form & der.      & bounded & needed \\
\hline 
This work & $\mu$ & $P(\rho, \epsilon)$ & No & Yes & No \\
Noble \cite{Noble:2006:626}
& $(D/\mu, v^2)$  & $P(\rho, h)$ & Yes & No & Yes \\
Siegel \cite{Siegel:2018:71}
& $ (D/\mu, T)$ & $P(\rho, T)$ & Yes & No & Yes \\
Duran \cite{Duran:2008:937} 
& $(W,D/\mu, T)$ & $\epsilon(\rho,T)$ &Yes & No & Yes \\
Neilsen \cite{Neilsen:2014:104029,Palenzuela:2015:044045}
& $D/\mu$ & $P(\rho, \epsilon)$ & No & Yes & No\\
Newman \cite{Newman:2014}
& $P$ & $P(\rho, h)$ & No & Yes & No 
\end{tabular} 
\end{table}

%===========================================================================
\section{Impact of Numerical Error in Evolved Variables} 
\label{sec:evol_err}

In the following, we investigate consequences of numerical errors in the evolved
variables in conjunction with the corrections of invalid states described in
\Sref{sec:enforce}. Further, we identify regions in parameter space where
the primitive variables are particularly sensitive  to errors of the evolved 
ones.

\subsection{Newtonian Limit}
\label{sec:newtonian}
It is instructive to consider the relation between evolved and primitive variables 
in the Newtonian limit.
Assuming that both kinetic and thermal specific energies are nonrelativistic 
corresponds to $v \ll 1, \epsilon \ll 1, a \ll 1, h\approx 1$ (for simplicity
we chose $m_B$ such that $h_0=1$ in this subsection). 
To leading order in $v^2$ and $\epsilon$, we obtain
\begin{align}
x &\to x_\mathrm{N} = \frac{1}{1+b^2}\\
\Fr^i &\to x_\mathrm{N} \Tro^i + \Trp^i \\
v^i &\to \Fr^i   \\
\epsilon &\to q - \frac{1}{2} \left( b^2 \left(1+v_\perp^2 \right) + v^2 \right)
\end{align}
Taking the Newtonian limit locally does not imply small $b$.
However, if the magnetic field energy is comparable to the rest mass density one 
cannot expect the velocity to stay non-relativistic during the course of the evolution. 
It is a plausible assumption that the density of kinetic energy is not much smaller than
the magnetic energy density. Setting $\mathcal{O}(b^2) \approx \mathcal{O}(v^2)$,
we find $x_\mathrm{N} \approx 1$.  Since $\mathcal{O}(E) = \mathcal{O}(vB)$, 
we can also neglect the electric contribution $b^2 v_\perp^2$ to $\epsilon$.

On a side note, it is easy to show that the master function \Eref{eq:master} 
becomes a linear function in the Newtonian limit (with $b^2\ll 1$). As 
$x(\mu) \approx 1$, $\Fr(\mu)$ and $\Fq(\mu)$ are independent of $\mu$ 
and equal to the correct values. The same holds in turn for
$\hat{\rho}$, $\hat{\epsilon}$, $\hat{P}$. 
Further, $\hat{\nu} \approx \hat{h} \approx h_0$. Since $\Fr^2 \ll 1$, 
the master function becomes $f(\mu) \approx \mu - 1 $.

We now turn to the propagation of the evolution error of the variables $\Tq,\Tr,D$. 
Even in the Newtonian limit, both $v^2$ and $b^2$ contributions can dominate 
$q$ if $\epsilon$ is even smaller. Since $v$ is essentially computed from $\Tr$, 
computing $\epsilon$ from the evolved variables suffers from cancellations 
that amplify the evolution errors. In detail,
\begin{align}
\begin{split}
\frac{\delta\epsilon}{\epsilon} 
=& \frac{\delta q}{q} \left( \mathcal{O}\left( \frac{b^2}{\epsilon} \right)
                             +\mathcal{O}\left( \frac{v^2}{\epsilon} \right)
                      \right) \\
&+ \frac{\delta b}{b} \mathcal{O}\left( \frac{b^2}{\epsilon} \right)
 + \frac{|\delta \Tr^i|}{\Tr} \mathcal{O}\left( \frac{v^2}{\epsilon} \right)
\end{split}
\end{align}

Once $\delta\epsilon/\epsilon$ exceeds unity, reconstructing $\epsilon$
from the evolved variables might lead to larger errors than simply setting 
it to the zero-temperature value.
Assuming some bound for the relative errors of the evolved variables,
this corresponds to critical values for $b^2$ and $v^2$. 

\subsection{Magnetically Dominated Regime}

In the context of magnetohydrodynamic evolution, magnetically dominated refers
to the magnetic pressure exceeding the fluid pressure. Increasing the
field strength at fixed matter density, the movement of matter becomes 
constrained along the field lines at some point.

This effect is also reflected in the equations we use for primitive recovery.  
The relation between total and fluid momentum $S_\perp$ components 
perpendicular to the magnetic field is $\Fcmom_\perp = x \Tcmom_\perp$,
as seen from \Eref{eq:rfvec}. The quantity $x$ depends only on $\mu b^2$. 
In the limit $\mu b^2 \gg 1$, we find $x \ll 1$.
In that case, the perpendicular part of the evolved momentum is dominated
by the electromagnetic part. However, the latter is proportional to
$v_\perp$ in ideal MHD, and also points in the same direction.
Therefore, the evolution error of the perpendicular part of momentum 
is not amplified by cancellations when recovering the fluid velocity
orthogonal to the field. Also the parallel part of the evolved momentum 
is not problematic since it has no electromagnetic contribution.

As discussed for the Newtonian limit, strong magnetic fields are also detrimental 
to the accuracy of $\epsilon$ since evolution errors in $B$ are amplified by 
cancellation. From \Eref{eq:hatq}, we find that the magnetic field contribution to the energy
causes strong cancellation error if $\Fq \ll b^2 (1-v_\perp^2)$.

\subsection{General case}

\begin{figure}
  \begin{center}
    \includegraphics[width=0.95\columnwidth]{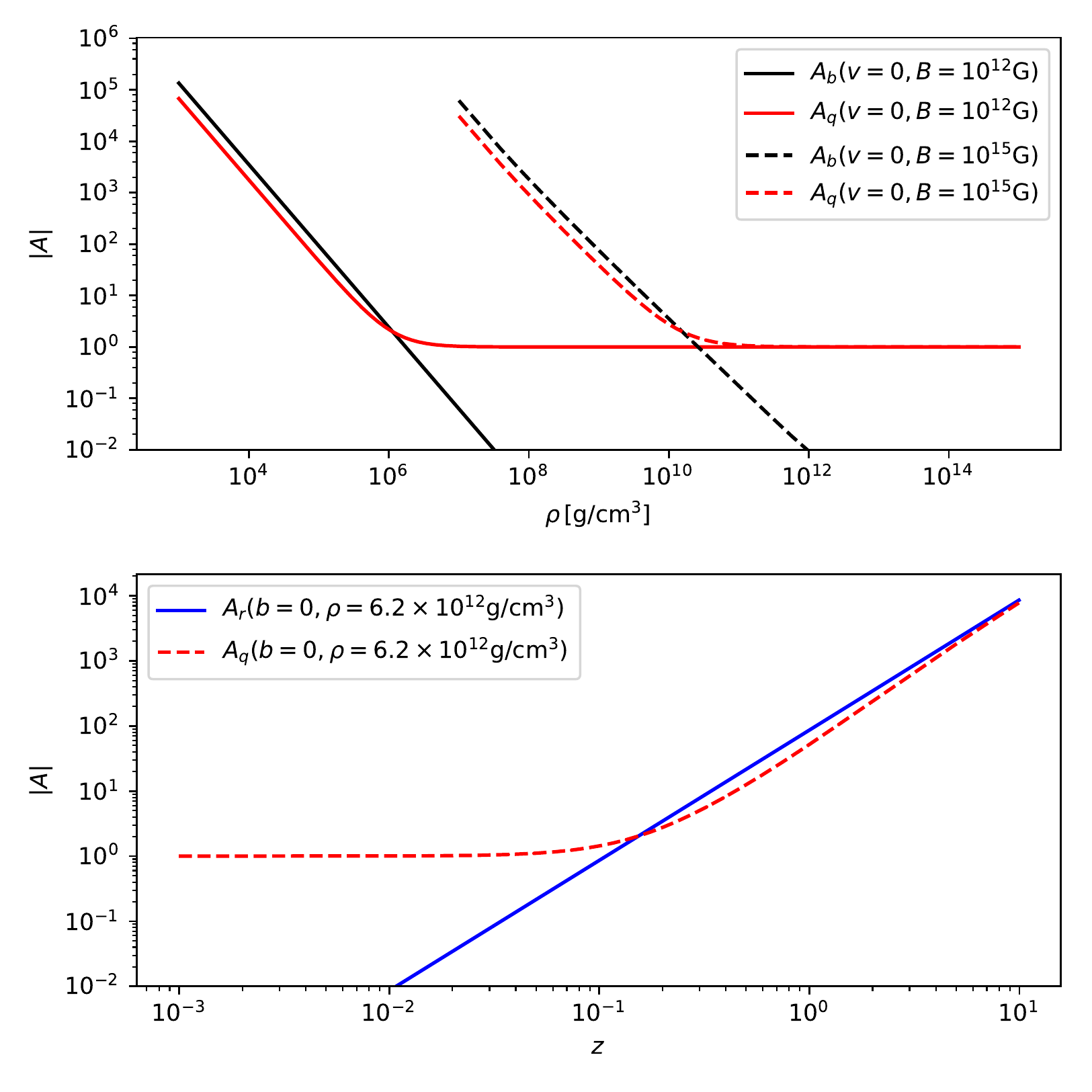}  
    \caption{Amplification factors $A_q$, $A_r$, and $A_b$ for the relative error 
    of $\epsilon$, as defined in \Eref{eq:amplfac},
    for the example of cold matter obeying the MS1 EOS.
    Top panel: amplification as a function of density, for fixed magnetic field
    and velocity.
    Bottom panel: amplification versus velocity, for fixed density and 
    no magnetic field.
 }
    \label{fig:error_ampl}
  \end{center}
\end{figure}

To quantify the error amplification in the general case, we compute the partial derivatives
of the primitive variables with respect to the conserved ones (by means of finite 
differences). 
As expected, specific internal energy and pressure exhibit large error amplification
in some regimes, while the fluid momentum is well behaved.
\Fref{fig:error_ampl} shows the behavior of amplification factors
of the specific energy error in relation to errors in the evolution
of $q$, $b$, and $r$, defined as
\begin{align}\label{eq:amplfac}
A_q &= \left.\frac{\delta\log(\epsilon)}{\delta\log(q)} \right|_{D,r,b} \\
A_b &= \left. \frac{\delta\log(\epsilon)}{\delta\log(b)} \right|_{D,q,r} \\
A_r &= \left. \frac{\delta\log(\epsilon)}{\delta\log(r)} \right|_{D,q,b}
\end{align}
The error of the pressure shows the same qualitative behavior.
For a magnetar-strength field $B=10^{15}\usk\mathrm{G}$,
we find that a relative evolution error of $\delta b/b = 10^{-4}$ would start to 
dominate the evolution of $\epsilon$ 
at densities of magnitude $10^{8} \usk\gram\per\centi\meter\cubed$. 
The same holds for a relative error $\delta q/q = 10^{-4}$, which 
is to be expected since $q$ is dominated by the $b^2$ contribution.  
This regime could be relevant
for the engine of SGRBs, as a popular model assumes a low-density 
funnel along the rotation axis of a black hole immersed in a strong magnetic 
field which is anchored in a surrounding disk. Similarly, the material surrounding a
supramassive (i.e.~long-lived) neutron star merger remnant could be affected.

The consequence of artificial heating could be artificial outflows and increased neutrino
luminosity. To assess 
the potential for spurious winds, we can compare the scales of additional 
specific energy caused by the evolution error and the specific gravitational 
binding energy. At a distance $100\usk\kilo\meter$ to a $M=2 M_\odot$ remnant, 
we find that the thermal error starts to dominate at densities 
$10^{8} \usk\gram\per\centi\meter\cubed$,
again for a fiducial evolution error $\delta b/b = 10^{-4}$ and $B=10^{15}\usk\mathrm{G}$.

On the other hand, the above discussion is overly pessimistic if the increase 
in thermal energy by physical effects, such as shocks or neutrino absorption, 
greatly exceeds the one by numerical errors. In other words, the presence of a 
mild outflow caused by mild heating should be met with more skepticism than
stronger outflows caused by prominent heating.

\subsection{Interaction with recovery corrections}
\label{sec:random}
Enforcing the evolved variables to stay in the physically
valid regime corresponds to a projection onto the validity 
boundary. Depending on the choice of projection, it is possible
that the corrections cause a drift along the boundary, in the 
worst case with a preferred direction. This is of particular 
concern for the very frequent correction of limiting the specific 
energy above the zero temperature value. For our scheme,
only the energy density is corrected in that case. Therefore,
it does not introduce a drift of the evolved momentum density.

The main effect of limiting $\epsilon$ above the zero temperature 
value may be to induce a spurious heating. The reason is that cutting the 
evolution error distribution from below creates a positive bias 
until the error distribution has little support below the cut. 
Of course, the raw error distribution before the correction could already contain 
a bias. For the idealized case in which the evolution errors follow a zero-mean 
normal distribution around the correct result, we expect the temperature
to increase until the thermal energy reaches a level comparable to the 
width of the error distribution for total energy. 

Note that excessive artificial heating could reduce the velocity, 
since the momentum density incorporates a factor $h$.
However, if $h$ is significantly increased non-homogeneously by the errors, 
the corresponding changes in thermal pressure can be expected to cause 
gradients and corresponding acceleration.

In the above discussion, we omitted the effect of finite root solving 
accuracy. Our implementation of the algorithm recomputes
all conserved variables from the primitive ones if, and only if, 
corrections were required. 
Therefore, the momentum only remains 
constant to the accuracy of the root solving when applying the correction
to the energy. This error is formally bounded by \Eref{eq:acc_cons}.
We cannot predict, however, if the distribution limited by this bound
is symmetric or not. Any bias might lead to a cumulative effect over
many corrections. For the worst possible scenario where each correction leads
to the maximum possible momentum error always pointing along the momentum,
a few thousand corrections could add up to intolerable levels.

To assess the actual behavior, we performed a numerical experiment. 
Instead of using a full numerical evolution, we employ 
a random walk model representing an evolution error, 
starting at selected states.
After each ``evolution'' step,
we apply the primitive recovery and limit the conserved 
variables to the allowed regime. The 
cumulative corrections applied to energy and momentum
are monitored.

In this approach, we can prescribe the error distribution. 
As a worst case example, we use a normal distribution
with negative mean for the energy error. Starting at a zero temperature 
state, this causes frequent corrections to the energy.
Note that the expected error in the momentum does not depend on the magnitude
of the corrections, but this randomized test nevertheless involves different
magnitudes. For the root finding accuracy, we use four different 
values $\Delta=10^{-7}, 10^{-8}, 10^{-9}, 10^{-10}$.

We find that the average momentum error is orders of magnitude smaller than 
the limit \Eref{eq:acc_cons}.
Selecting an initial state $v=0.99, b=2, \rho=6\times 10^{12} \usk\gram\per\centi\meter\cubed$,  
the momentum errors of individual correction steps approach machine precision levels 
around $\Delta=10^{-9}$.
We believe that the reason might be that the 
accuracy increases drastically during the final root finding step, 
such that the average root error is much
smaller than the prescribed maximum. We conclude that cumulative effects
of the corrections can likely be neglected. In case of evidence to the contrary,
the solution would be to simply not recompute the momentum, sacrificing 
machine-precision consistency for error reduction.

We also apply the random walk model to states with different combinations of
$b=\{0,2\}$, $v=\{0, 0.99\}$, $\epsilon_\mathrm{th} = \{0,10\}$, perturbing the evolved variables 
separately with normally distributed relative errors of order $10^{-4}$.
This test confirms that the implementation of the zero-temperature energy 
correction works as intended. 
We monitored the behavior of  $v^i, W, \epsilon, P, \mu$  for the above cases 
and did not encounter any problematic behavior.

%===========================================================================
\section{Summary and Discussion}
\label{sec:summary}

In this work, we solved the technical problem of primitive variable recovery
in relativistic ideal magnetohydrodynamic evolution codes via a new fully reliable scheme.
We derived a mathematical proof that the algorithm always finds a valid solution,
and that the solution is unique.
Moreover, we derived expressions that allow us to prescribe the accuracy of the 
individual primitive variables. 

The guaranteed reliability of the new algorithm is a big advantage compared
to older methods, which are able to handle most of the parameter space encountered
in BNS merger scenarios, but may still fail in some 
cases \cite{Siegel:2018:71}.
Even rare recovery failures are very problematic, since they necessitate  
manual intervention, and may require repeating parts of the simulation.
Recovery failures are practically unpredictable and potentially chaotic (we recall the
Newton-Raphson fractal related to convergence properties of a standard root finding procedure).
This is aggravated for recovery schemes that rely on an initial guess taken from the 
previous timestep. 
The automated approach of using a fixed state (e.g., artificial atmosphere) in case of 
recovery failure will render simulations unpredictable in practice. 

The ability to identify unphysical evolved variables, as well as the nature of
the invalidity, is another advantage of our method.
All evolution schemes produce unphysical states occasionally, most of which 
are harmless. However, sometimes invalid input occurs as the first symptom of more severe 
evolution errors. 
Our method allows to prescribe an error policy and selectively apply
corrections based on the nature of the problem. 
Such corrections (or lack of corrections) should be considered as part of 
the evolution scheme, but they are mentioned in the literature only rarely.

The design of our scheme naturally suggests a particular prescription 
for correcting slightly unphysical input. 
We discussed potential cumulative effects of those corrections, and predicted
that it will create artificial heating if the matter is close to zero temperature.
We also showed that there should be no direct impact on the momentum.
We validated this by performing a numerical experiment using random walk perturbations
to emulate evolution errors.

Since the implementation of recovery algorithms is a work-intensive endeavor,
we are making our reference implementation public in form of a well-documented 
library named \texttt{RePrimAnd} \cite{RePrimAnd}, which can be used by any evolution code.
In order to be useful in practice, the recovery should not constrain the type
of EOS. Therefore, our recovery algorithm is formulated in an EOS-agnostic 
manner, and the reference implementation contains a generic interface 
for using arbitrary EOS.

We subjected the code to a comprehensive suite of tests, demonstrating 
that both the algorithm and the actual implementation are robust up to  
Lorentz factors and values of magnetization much larger than those relevant for
BNS mergers. We also showed that the scheme is computationally 
efficient regarding the number of EOS evaluations (efficiency of
EOS implementations aside).

While investigating the accuracy of the recovery scheme, we identified regimes
where rounding errors are amplified by unavoidable cancellation errors. 
We quantified the dominant contributions and found that the accuracy measured 
in our tests is compatible with the predictions. We also found that the rounding 
errors are irrelevant because the very same cancellation also leads to the 
amplification of evolution errors. Investigating the error propagation from 
evolved to primitive variables, we showed that the accuracy of the thermal 
energy and thermal pressure can severely degrade when evolving strongly magnetized 
regions of low density. 

We believe that our results will be useful in particular for studying
the launching mechanism of jets powering SGRBs, 
as well as the mass ejection processes that are ultimately responsible for kilonova signals.
Both astrophysical scenarios involve strongly magnetized matter.

%===========================================================================
\acknowledgments

\noindent This work was supported by the Max Planck Society's Independent 
Research Group Programme.
J.\,V.\,K. kindly acknowledges the CARIPARO Foundation for funding his Ph.D. fellowship within the Ph.D. School in Physics at the University of Padova.

%===========================================================================
%\section*{References}
\bibliographystyle{apsrev4-1-noeprint}
\bibliography{article}

\appendix

\section{Derivations}
\label{app:deriv}
In this appendix, we provide derivation steps left out in the main
text. 
First, we derive \Eref{eq:masterderiv} for the derivative of the master function.
Starting from \Eref{eq:master}, 
\begin{align}
f &= \mu - \hat{\mu} = \mu - \left( \hat{\nu} + \Fr^2 \mu \right)^{-1} \, , \\  
f' &= 1 + \left( \hat{\nu} + \Fr^2 \mu \right)^{-2} 
           \left( \hat{\nu}' + \Fr^2 + 2 \Fr \Fr' \mu  \right) \, , \\
 &= 1 + \hat{\mu}^2  \left( \hat{\nu}' + \Fr^2 - 2 \left(1-x\right)x^2 \Tro^2 \right) \, , \label{eq:dfdmuraw}
\end{align}
where primes denote derivatives with respect to $\mu$.
We first consider the case where $\epsilon$ computed from \Eref{eq:hateps} does not
exceed the upper limit allowed by the EOS (but may be smaller than the zero temperature 
limit). For those cases, we can use \Eref{eq:hatnuderiv} to get
\begin{align}
\begin{split}
f' &= 1 + \hat{\mu}^2  \Biggl( -\left(1+c_s^2\right) \hat{\nu} \frac{\hat{W}'}{\hat{W}} \\
      &\qquad\qquad\quad  + \Fr^2 - 2 \left(1-x\right)x^2 \Tro^2 \Biggr)
\end{split}\\
\begin{split}
&= 1 + \hat{\mu}^2  \Biggl( -\left(1+c_s^2\right) \hat{\nu} \hat{W}^2 \mu \left( x^3 \Tro^2 + \Trp^2 \right) \\
   &\qquad\qquad\quad + x^2 \Tro^2 + \Trp^2 - 2 \left(1-x\right)x^2 \Tro^2 \Biggr)
\end{split}
\end{align}
At a solution, we have $\hat{\mu}=\mu$ and  $\hat{\nu} \hat{W}^2 \mu = 1$, which leads to 
\begin{align}
\begin{split}
f' &= 1 + \mu^2  \Biggl( -\left(1+c_s^2\right) \left( x^3 \Tro^2 + \Trp^2 \right) \\
   &\qquad\qquad\quad + x^2 \Tro^2 + \Trp^2 - 2 \left(1-x\right)x^2 \Tro^2 \Biggr)
\end{split} \\
  &= 1 + \mu^2  \left( \left(1 - c_s^2\right) \left( x^3 \Tro^2 + \Trp^2 \right) 
         -x^2 \Tro^2 - \Trp^2  \right) \\
  &= 1 - \mu^2 \Fr^2 + \mu^2  \left(1 - c_s^2\right) \left( x^3 \Tro^2 + \Trp^2 \right) 
\end{align}
Using that $\hat{v} = \mu \Fr$ at the solution, we arrive at \Eref{eq:masterderiv}.

We now address the case where $\epsilon$ computed from \Eref{eq:hateps} does exceed the 
limit $\epsilon_\mathrm{max}$ below which the EOS is valid. \Eref{eq:nuhat} then becomes
\begin{align}
\hat{\nu}  &= \nu_B 
= \left(1 + \hat{a}\right)\left(1 + \Fq - \mu \Fr^2 \right) \\
\hat{\nu}'  
  &= \frac{\hat{\nu}}{1+\hat{a}}\hat{a}'  
     + \left(1+\hat{a}\right) \left( \Fq' - \Fr^2 -2\mu\Fr\Fr'\right)
\end{align}
Inserting \Eref{eq:drbardmu}, \Eref{eq:dqbardmu}, and \Eref{eq:rbsqr} yields
\begin{align}
\frac{\hat{\nu}'}{\hat{\nu}}
&= \frac{\hat{a}'}{1+\hat{a}} - \frac{1+\hat{a}}{\hat{\nu}} R^2
\end{align}
with $R^2 \equiv x^3 \Tro^2 + \Trp^2 \le \Fr^2$.
For the case at hand, 
${\hat{a}(\hat{\rho}) = a(\hat{\rho}, \epsilon_\mathrm{max}(\hat{\rho}))}$, 
and hence
\begin{align}
\hat{a}' &= \hat{\rho}' \left( \frac{\partial a}{\partial \rho} 
               + \frac{\mathrm{d}\epsilon_\mathrm{max}}{\mathrm{d}\rho}\frac{\partial a}{\partial\epsilon}  \right)
\end{align}
Splitting the derivative of $\epsilon_\mathrm{max}$ into adiabatic
and residual contributions by using definition \Eref{eq:defadiabA}, we obtain
\begin{align}
\hat{a}' &= \hat{\rho}' \left(\frac{\partial a}{\partial \rho} 
              + \left(\frac{\hat{P}}{\hat{\rho}^2} 
                        + \frac{A(\hat{\rho})}{\hat{\rho}} \right) 
                                     \frac{\partial a}{\partial\epsilon} \right) \\
  &= \hat{\rho}' \left( \left.\frac{\mathrm{d} a}{\mathrm{d} \rho}\right|_s 
              + \frac{A(\hat{\rho})}{\hat{\rho}}  
                                     \frac{\partial a}{\partial\epsilon} \right)                                      
\end{align}
One can express the adiabatic soundspeed in terms of $a$ as
\begin{align}
\left.\frac{\mathrm{d} a}{\mathrm{d} \rho}\right|_s
&= \frac{1+a}{\rho} \left( c_s^2 - a \right)
\end{align}
which allows us to write
\begin{align}
\hat{a}' 
&= \frac{\hat{\rho}'}{\hat{\rho}} 
       \left( \left(1+\hat{a}\right) \left(c_s^2 - \hat{a}\right) 
       + A \frac{\partial a}{\partial\epsilon} \right)
\end{align}
Evaluating at the solution, we can use $\hat{\mu}=\mu$ and $\hat{W}^2\hat{\nu}\mu=1$.
Using also \Eref{eq:dhatwdmu} and \Eref{eq:dhatrhodmu}, we get
\begin{align}
\frac{\hat{\nu}'}{\hat{\nu}}
&= -\hat{W}^2 \mu R^2
  \left(1 + c_s^2 + \frac{A}{1+ \hat{a}}\frac{\partial a}{\partial\epsilon} \right)
\end{align}
Inserting into \Eref{eq:dfdmuraw}, we can rewrite
the master function derivative at the solution as
\begin{align}
f' &= 1 - \hat{v}^2 + \hat{v}^2 \frac{R^2}{\Fr^2} \left(1 - c_s^2 
                - \frac{A}{1+ \hat{a}}\frac{\partial a}{\partial\epsilon} \right) 
\end{align}
If \Eref{eq:req_epsmax} holds, $f'$ is always strictly positive at the solution, 
as claimed in \Sref{sec:uniqe}.

Finally, we need to discuss the corner case where the mass density 
computed in \Eref{eq:hatrho} is outside the valid range of the EOS.
If 
$\Fcrmd / \hat{W}(\mu_+) > \rho_\mathrm{max}$ or
$\Fcrmd < \rho_\mathrm{min}$, there is no solution
with valid density and no need to determine the root. 
The only complication arises when 
$\Fcrmd / \hat{W}(\mu_+) <  \rho_\mathrm{max} < \Fcrmd$ or 
$\Fcrmd / \hat{W}(\mu_+) <  \rho_\mathrm{min} < \Fcrmd$.
Although the proof for the existence of a solution remains valid in those 
cases, we did not succeed to prove uniqueness of the master function root 
on the range $(0,\mu_+]$. 
Luckily, it is not necessary to prove uniqueness on the full interval.
Instead, we numerically solve $\Fcrmd / \hat{W}(\mu) = \rho_\mathrm{min/max}$ 
for $\mu$. We recall that $\hat{W}(\mu)$ is an analytic expression that 
does not involve the EOS, and monotonically increases with $\mu$.
Hence we can find a sub-interval $(\mu_a, \mu_b) \subseteq (0,\mu_+)$
that consists of all values for which the density $\Fcrmd/\hat{W}(\mu)$
is valid.
We already know that the master function $f$ has at most one root 
on that interval. There is a solution with valid density if and only if
$f(\mu_a)$ and $f(\mu_b)$ have opposite sign, which is easy to check.
Using an initial bracket $(\mu_a, \mu_b)$ for the root finding then ensures
uniqueness.

\end{document}